\documentclass[twocolumn,aps,nofootinbib,amsfonts,superscriptaddress]{revtex4}
\usepackage{amsfonts,amssymb,amscd,amsmath}
\usepackage{graphicx}

 \newcommand\beq{\begin{equation}}
 \newcommand\eeq{\end{equation}}
 \newcommand\beqn{\begin{eqnarray}}
 \newcommand\eeqn{\end{eqnarray}}

\def\fm{\,\mbox{fm}}
\def\GeV{\,\mbox{GeV}}

\def\fm{\,\mbox{fm}}
\def\GeV{\,\mbox{GeV}}

\def\eq#1{{Eq.~(\ref{#1})}}

\begin{document}

\title{Hadrons and direct photon in pp and pA collisions at LHC and saturation effects }

\author{Amir H. Rezaeian}
\affiliation{Institut f\"ur Theoretische Physik, Universit\"at Regensburg, 93040 Regensburg, Germany}
\affiliation{Departamento de F\'\i sica y Centro de Estudios
Subat\'omicos,\\ Universidad T\'ecnica
Federico Santa Mar\'\i a, Casilla 110-V, Valpara\'\i so, Chile}
\author{Andreas Sch\"afer }
\affiliation{Institut f\"ur Theoretische Physik, Universit\"at Regensburg, 93040 Regensburg, Germany}

\begin{abstract}
 We investigate hadrons and direct photon production in $pp$ and $pA$
 collisions at the energies of RHIC and LHC within the color-dipole
 approach employing various saturation models. We show that greatest sensitivity to
 saturation effects is reached at very forward rapidities for $pp$
 collisions at LHC ($\sqrt{s}=14$ TeV).  The ratio of direct-photon to pion $\gamma/\pi^0$ production can be about
 $20 \div 10$ (at $\eta=7\div 8$ ). Therefore, direct
 photon production at forward rapidities should provide a rather
 clean probe. We calculate the rapidity dependence of the
 invariant cross-section and find some peculiar enhancement at
 forward rapidities which is more pronounced for direct
 photon production. We show that this peak is further enhanced by saturation effects. We provide
 predictions for the nuclear modification factor $R_{pA}$ for pions
 and direct photon production in $pA$ collisions at LHC energy at
 midrapidity. We show within various saturation models that the pion Cronin enhancement at RHIC is
 replaced by a moderate suppression at LHC energy at midrapidity
 due to gluon shadowing effects. Cronin enhancement  of direct photons can survive at LHC energy within
 models with a larger saturation scale.

\end{abstract}

\maketitle
\date{\today}

\section{Introduction}

The Large Hadron Collider (LHC) will allow to explore a new regime of QCD where parton saturation
effects become important \cite{Gribov:1984tu1,Gribov:1984tu,Jalilian-Marian:1997jx,Kovchegov:1999yj,Iancu:2003xm}. 
At the same time, the physics of saturation might also be relevant for a detailed understanding of the underlying events, i.e. the backgrounds for New Physics searches at LHC.

It is believed that $pp$ and $pA$ collisions provide a testing ground
to disentangle the initial- and final-state effects in $AA$ collisions and can be
used as a baseline for understanding the physics of heavy-ions collisions.
For example, to interpret jet-quenching, a precise
and firm understanding of the Cronin, shadowing and saturation effects in $pA$ collisions is indispensable.

The Color Glass Condensate (saturation) approach to QCD at high energy \cite{Gribov:1984tu1,Gribov:1984tu,Jalilian-Marian:1997jx,Kovchegov:1999yj,Iancu:2003xm}
has been very successful to describe a variety of processes at Relativistic Heavy Ion Collider
(RHIC) \cite{rhic-s}(for a review see \cite{Iancu:2003xm} and references
therein). Nevertheless, the importance of saturation effects 
is still disputable given that other approaches offered alternative
descriptions, see for example Refs.~\cite{break,fin1}. In order to test saturation physics and its relevance,
it seems therefore mandatory to consider various reactions in different
 kinematic regions at LHC and future collider
experiments. 

Here, we study hadron and direct
photon production in $pp$ and $pA$ collisions within the
light-cone color-dipole formulation and investigate the role of
saturation and shadowing at LHC energies. 
The corresponding phenomenology is based on the universal
$q\bar{q}$ dipole cross-section.  The dipole cross-section incorporates the multiple gluon
scattering and non-linear gluon recombination effects and can be
in principle measured in deep-inelastic scattering (DIS), see section VI. In the parton model language,  the dipole
cross-section plays the role of leading twist parton distributions in
an all twist environment.

Direct photons (photons radiated in hadronic collisions not via
hadronic decays) carry important information about the collision
dynamics which is undisturbed by final state interactions.  We compare hadron and direct
photon production mechanisms at various energies and rapidities in $pp$ collisions. We show that the ratio of
photon/pion production at very forward rapidities grows and can become as big as
one order of magnitude at the LHC energy  $\sqrt{s}=14$ TeV. Measurements of direct photons at forward rapidities should be rather clean, as
the background from radiative hadronic decays is significantly 
suppressed\footnote{Experimentally measurements at forward rapidities are a challenge since production rates are lower due to kinematic limits.}. At the same time, we show that both hadrons and direct
photons are sensitive to saturation effects at forward rapidities
at $\sqrt{s}=14$ TeV $pp$ collisions.

We also investigate the role of saturation and shadowing effects for
hadron and direct photon production in $pA$ collisions at LHC. Our
approach gives a rather fair description of PHENIX data for the Cronin
ratio $R_{pA}$ of pions. We show that the nuclear modification
factor $R_{pA}$ for $\pi^0$ at LHC ($\sqrt{s}=5.5 $ TeV) at
midrapidity becomes less than $1$ in all saturation
color-dipole models due to gluon shadowing. The suppression obtained
(for $R_{pA}$) in our approach is less than the one predicted in the
Color Glass Condensate (CGC) approach \cite{CGC-cronin}. We will later
highlight the difference between our results and other reported
predictions. We will also show that the nuclear modification factor $R_{pA}$ for direct photons
is also less than $1$ within the CGC color-dipole model once shadowing effects are included. In contrast, the Cronin enhancement for
photons can survive even after inclusion of shadowing effects within
the Golec-Biernat and W\"usthoff color-dipole model which has a bigger
saturation scale than the CGC model.

The paper is organized as follows: In Sec. II and III we calculate
gluon radiation from projectile gluons and quarks in the color-dipole
approach. In Sec. IV we introduce the light-cone color-dipole
factorization scheme for hadron production. 
In Sec III, IV we will also highlight the differences between our approach
with others. In Sec V we calculate the
direct-photon production in $qN$ and $pp(A)$ collisions. In Sec VI we
introduce gluon saturation within various approaches and color-dipole
models. In Sec VIII we discuss nuclear gluon shadowing, Cronin effect
and nuclear modification factor for partons, pions and direct photon
production.  In Sec VII and VIII we present our numerical results for
both hadron and direct-photon production in $pp$ and $pA$ collisions,
respectively. As a conclusion, in Sec. IX we highlight our main
results and predictions for LHC.

\section{gluon radiation by a projectile gluon: $gN(A)\to g_1g_2X$} 
The underlying mechanisms of the multiple particle interactions is
controlled by the coherence length $l_c$.  In the incoherent case, the
multiple interaction amplitude can be simplified as convolution of
differential cross sections while in the coherent case, one should
convolute scattering amplitudes rather than differential
cross-sections.

The coherence length $l_c$ can be estimated from 
the inverse longitudinal momentum transfer, 

\beq l_c\equiv \frac{2E_i}{M^2}\equiv \frac{2E_i\alpha(1-\alpha)}{k_{T}^2},
\label{g-1g} 
\eeq 
where $E_i$ is the initial parton energy and $k_T$
is the relative transverse momentum of the final partons. In the above equation, $M$ is the invariant mass of the two final partons, neglecting parton masses.  The parameter $\alpha$
is the fractional light-cone momentum of one of the final partons.
Gluon radiation is dominated by small values of $\alpha\ll1$, therefore we have, 
\beq l_c\approx \frac{2E_f}{k_{T}^2}\approx \frac{\langle z \rangle  \sqrt{s}}{m_N p_T},
\label{g-2g} 
\eeq 
where $E_f$ is the energy of the parton detected in the final state, $p_T$ is the transverse momentum of the fragmented
hadron at midrapidity, and $m_N$ is the nucleon mass. For pion
production, the average momentum fraction $\langle z \rangle$ in the
fragmentation functions is about $0.4-0.6$ in the range of $2\leq p_T~(\text{GeV})\leq 8$.
For a coherence length which is shorter than the typical internucleon
separation $l_c \lesssim R_A$ (where $R_A$ denotes the nuclear
radius), the projectile interacts incoherently. At the RHIC energy
$\sqrt{s}=200\GeV$ and intermediate $p_T$ we are almost in the
transition region between the short- and long-coherence length regime. 
In more central collisions, at higher $p_T$ we are in the
short-coherence length (SCL) limit and at LHC energies at moderate $p_T$ we are again in the long-coherence length (LCL) limit.

\begin{figure}[!t]
       \includegraphics[width=9 cm] {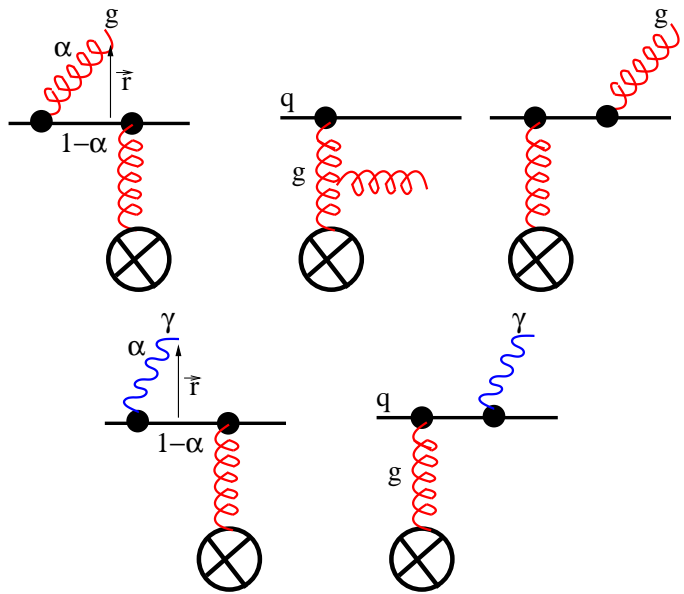}
       \caption{Gluon radiation (top panel) and direct photon production (bottom panel) for a projectile quark interacting with the target. }\label{f1}
\end{figure}

There is much experimental evidence for a large intrinsic momentum of gluons, see Refs.~\cite{kst2,spot} and reference therein. Therefore, interaction with spectators
is important since color screening is at work. At smaller and moderate
$p_T$ one should then include interaction with spectators,
i.e. instead of "elastic" gluon scattering, $gN\to gX$, we need to
consider bremsstrahlung subprocesses, $gN\to ggX$, or $qN\to qgX$. The
lowest order for these processes includes the three graphs shown in Fig.~\ref{f1}
(interactions with the initial and two final partons).  After summing
over radiated gluons, the cross section of this reaction can be
expressed in terms of the color-dipole amplitudes \cite{kst2,kst1},
and can be diagonalized for a nuclear target provided that the coherence
length is sufficiently long. Note that since parton trajectories before and after gluon (or photon) radiation have
different impact parameters, and the corresponding terms in the
bremsstrahlung amplitude have different signs, one arrives at an
expression, which is formally identical to the amplitude of an
inelastic dipole-target interaction.  This is only a formal
procedure of calculation, while no real $q\bar{q}$ color-dipole is involved in the
process of radiation contrary to DIS where a photon does split into a real $q\bar{q}$ pair.

In the LCL regime, the transverse momentum spectra of gluon
bremsstrahlung for a high energy gluon interacting with a nucleon N
(or nucleus A) including the nonperturbative interactions of the
radiated gluon reads \cite{kst2,kst1},
\beqn
&&\frac{d\sigma^{gN(A)\to g_1g_2X}}{d^2\vec{k}_T}(k_T, x) = \frac{1}{(2\pi)^2}
\int d^2b~d^2r_1~d^2r_2 \nonumber\\
&\times&e^{i\vec k_T(\vec r_1-\vec r_2)}\overline{\Psi_{gg}^*(\vec r_1,\alpha )\Psi_{gg}(\vec r_2,\alpha )} \Bigl[
\mathcal{N}^{N(A)}_{3g}(\vec b,\vec r_{1},x)\nonumber\\
&+&
\mathcal{N}^{N(A)}_{3g}(\vec b,\vec r_{2},x)
- \mathcal{N}^{N(A)}_{3g}(\vec b,(\vec{r}_{1}-
\vec{r}_{2}), x)\Bigr],
\label{80}
 \eeqn
where $\alpha=p_{+}(g_1)/p_{+}(g)\ll 1$ denotes the light-cone momentum
fractional of the radiated gluon.  The partial amplitude
$\mathcal{N}_{3g}^{N}$ of a $3$-gluons system colliding with a proton at impact parameter $\vec b$ can be written in terms of the $q\bar{q}$ dipole
amplitude \cite{zakh,zakh1},
\beqn
 \mathcal{N}_{3g}^{N}(\vec{b},\vec{r},x)&=&\frac{9}{8}\Big\{ \mathcal{N}_{q\bar{q}}^{N}
 (\vec{b},\vec{r},x)+\mathcal{N}_{q\bar{q}}^{N}(\vec{b},\alpha\vec{r},x)\nonumber\\
&+&\mathcal{N}_{q\bar{q}}^{N}(\vec{b},(1-\alpha)\vec{r},x)\Big\},
\label{3g-1} 
\eeqn
 where the factor $9/8$ is the ratio of Casimir
factors. Here the vectors $\vec r$, $\alpha \vec r$ and
$(1-\alpha)\vec r$ denote the two gluon transverse separations $\vec
r (g_1)-\vec r (g_2)$, $\vec r (g)-\vec r (g_2)$ and $\vec r
(g)-\vec r (g_1)$, respectively.

Note that Eq. (\ref{3g-1}) can be simply understood by looking at several
limiting cases: if $r$ goes to zero, the transverse separation of final state
gluons $g_1$ and $g_2$ becomes zero leading to
$\mathcal{N}_{3g}^{N}(\vec{b},\vec{r}\to 0,x)=0 $ which reflects the fact
that a point like gluon-gluon fluctuation cannot be resolved by
interactions. In the two limiting cases of $\alpha\to 0, 1$, the
three-gluon system will be reduced to 
the two-gluon system which can be
then related to the $q\bar{q}$ dipole cross-section via 
the Casimir factor:
$\lim_{\alpha\to 0,1}\mathcal{N}_{3g}^{N}(\vec{b},\vec{r},x)=\frac{9}{4}\mathcal{N}^{N}_{\bar{q}q}(\vec
b,\vec{r},x)$ where $9/4$ is the ratio of the octet and triplet
color Casimir factor.  

The $q\bar{q}$ dipole amplitude in Eq.~(\ref{3g-1}) is related to the dipole-proton cross-section by integration over impact parameter, 
\beq
\sigma_{q\bar{q}}(r,x)=2\int d^{2}\vec b~\mathcal{N}_{q\bar{q}}^{N}(\vec b,\vec{r},x). 
\label{di-app}
\eeq

We still have to specify the light-cone distribution function
($\Psi_{gg}$) for the $gg$ Fock component fluctuations of the incoming
gluon, which includes nonperturbative interactions of these gluons.
 The light-cone wave function of the gluon-gluon (and quark-gluon) Fock
component of a gluon (quark) was calculated in Ref.~\cite{kst2} within a model
describing the nonperturbative interaction of gluons via
a phenomenological light-cone potential of an oscillatory form.
This is given by, 
\beqn 
&&\Psi_{gg}(\vec r,\alpha) =
\frac{\sqrt{8\alpha_s}}{\pi\,r^2}\,
\exp\left[-\frac{r^2}{2\,r_0^2}\right]\, \Bigl[\alpha(\vec
e_1^{\,*}\cdot\vec e)(\vec e_2^{\,*}\cdot\vec r) \nonumber\\
&+& (1-\alpha)(\vec
e_2^{\,*}\cdot\vec e)(\vec e_1^{\,*}\cdot\vec r) -
\alpha(1-\alpha)(\vec e_1^{\,*}\cdot\vec e_2^{\,*})(\vec e\cdot\vec
r) \Bigr],\ \label{100}\nonumber\\
 \eeqn
where $r_0=0.3\fm$ is the parameter characterizing the strength of
the nonperturbative interaction which has been fitted to data
on diffractive $pp$ scattering \cite{kst2}. In Eq.~(\ref{80}) the
product of the wave functions is averaged over the initial gluon
polarization, $\vec e$, and summed over the final ones, $\vec
e_{1,2}$.

 Based on pQCD one might expect that the gluon-gluon potential differs 
from the quark-antiquark one simply by a Casimir factor $9/4$. 
However, there exists
plenty of evidence indicating that the interaction of gluons is much
stronger due to non-trivial properties of the QCD vacuum, see Ref.~\cite{spot} and
references therein.  It turns out that the exact shape of the
light-cone gluon-gluon (quark-gluon) potential is not crucial \cite{kst2}. What is only important is the smallness of the mean
quark-gluon separation $r_0$ which defines the effective strength of gluons interaction. The value of $r_0=0.3$ fm obtained from
analysis of diffractive data \cite{kst2} agrees with both lattice calculations \cite{b1-n} and
also with the phenomenological model of the instanton liquid \cite{b2-n}.

We consider here the asymptotic expression of the gluon radiation
cross-section given in Eq.~(\ref{80}) for $\alpha\to 0$ which is
reliable at very long coherence lengths. This is certainly valid at
LHC energies. At RHIC energies, for hadrons produced at midrapidity
with moderate $p_T$, we are in the transition region between the
regimes of long and short coherence lengths. Moreover, the color-dipole
models we use in this paper, were fitted to DIS data at very small Bjorken-x
$x_B\leq 0.01$, which corresponds to $p_T\leq 2$ GeV at RHIC. Therefore, the prescription presented here should be less reliable at high-$p_T$
at RHIC energy. We will come back to this point in Sec. VIII. 

After some algebra one obtains,
\beqn 
&&\frac{d\sigma^{gN\to g_1g_2X}}{d^2\vec{k}_T\,d^2\vec{b}} = \frac{9\alpha_s}{\pi^3}
\int_0^{\infty}dr 
~\mathcal{N}_{q\bar{q}}^{N}(\vec{b},\vec{r})\nonumber\\
&\times&\Biggl\{ \frac{4\pi}{k_T}\left(1-e^{-k_T^2r_0^2/2}\right)J_1(k_T r) e^{\frac{-r^2}{2r_0^2}}\nonumber\\
&-&J_0(k_T r)e^{\frac{-r^2}{4r_0^2}}f(r)
 \Biggl\},
 \label{vn1}\
\eeqn 
where the function $f(r)$ is defined as
\begin{eqnarray}
f(r)&=& \int_{0}^{\infty} d\Delta \int_{-\pi}^{+\pi} d \theta \frac{(\Delta^2-r^2)
\Delta r}{(\Delta^2+r^2)^2-4(\Delta r\cos(\theta))^2}
e^{-\frac{\Delta^2}{4r_0^2}} \nonumber\\
&=&\pi r e^{r^{2}/4r_{0}^{2}}\left(Ei(\frac{-r^{2}}{4r_{0}^{2}})-2Ei(\frac{-r^{2}}{2r_{0}^{2}})\right).
\label{not0}\
\end{eqnarray}
In the case of a nuclear target the functional form of Eq.~(\ref{80}) still holds, but the
dipole amplitude for a nucleon target $\mathcal{N}_{3g}$ should be replaced
by the one for a nuclear target $\mathcal{N}^{A}_{3g}$.  
The partial elastic
amplitude $\mathcal{N}^{A}_{3g}$ for a colorless three-gluon system colliding
with a nucleus $A$ can be written in terms of the partial amplitude
$\mathcal{N}_{3g}^{N}$ of a three-gluon system colliding with a proton at
impact parameter $\vec b$, 
\beqn 
\mathcal{N}^{A}_{3g}(\vec{b},\vec{r},x)&=& 2
\Biggl\{1-
e^{-\int d^{2}\vec{s}~\mathcal{N}_{3g}^{N}(\vec{s},\vec{r},x)T_{A}(\vec{b}+\vec{s})}\Biggl\},\nonumber\\
\label{eik2} 
\eeqn 
where the $3$-gluons amplitude $\mathcal{N}_{3g}^{N}$ is related to the
$q\bar{q}$ dipole amplitude via Eq.~(\ref{3g-1}) and $ T_A(b)$ is the 
nuclear thickness function normalized to $\int d^2b T_A(b)=A$.  In a
very similar fashion as for the nucleon target case, one can analytically
carry out some of the integrals,
\beqn 
&&\frac{d\sigma^{gA\to g_1g_2X}}{d^2\vec{k}_T\,d^2\vec{b}} = \frac{4\alpha_s}{\pi^3}
\Big(\int_0^{\infty}dr 
\Biggl\{ -\frac{4\pi}{k_T}\left(1-e^{-k_T^2r_0^2/2}\right)\nonumber\\
&\times&J_1(k_T r) e^{\frac{-r^2}{2r_0^2}-\mathcal{I}_G(b, r)} 
+J_0(k_T r)e^{\frac{-r^2}{4r_0^2}-\mathcal{I}_G(b, r)}f(r)\Biggl\}\nonumber\\
&+&\frac{(2\pi)^2}{k_T^2}\left(1-e^{-k_T^2r_0^2/2}\right)^2\Big),\label{ga}\ 
\eeqn 

with the notation,
\begin{eqnarray}
\mathcal{I}_G(b, r)&=&
\frac{9}{4}\int d^{2}\vec{s}~\mathcal{N}_{q\bar{q}}^{N}(\vec{s},\vec{r})T_{A}(\vec{b}+\vec{s}),\nonumber\\
&\approx&\frac{9}{8}\sigma_{q\bar{q}}(r,x)T_A (b),
\label{not1}\
\end{eqnarray}
where in the second line we used Eq.~(\ref{di-app}) and ignored
possible correlations between the color-dipole amplitude and nuclear thickness. Notice that the second line were identically true if
the nuclear profile  would be a constant. In a more sophisticated approach
in order to properly incorporate the correlation between the
color-dipole amplitude and the nuclear thickness, one should also have
a model for the dipole amplitude which depends on the angle between
the dipole transverse radius $\vec r$ and the impact parameter $\vec
b$.  Unfortunately, with available HERA data, it is difficult to
incorporate the color-dipole orientation and most dipole models fitted
to HERA data depend only on the absolute value of the transverse dipole size
$|\vec r|$ and impact parameter $|\vec b|$. For a recent attempt to
incorporate the color dipole orientation, see
Ref.~\cite{mev2}. It has been
shown that the color-dipole orientation gives rise to azimuthal
asymmetries \cite{mev2}, but is unimportant for total cross-sections.

The remaining integrals
in Eqs.~(\ref{vn1},\ref{ga}) can be performed only numerically.


\section{gluons radiation by a projectile quark: $qN(A)\to qgX$} 
Gluon radiation of a projectile quark interacting with a
nucleon(nucleus) $qN(A)\to qgX$ can be calculated in a similar way as
outlined in the previous section. The cross-section is given by \cite{kst2,kst1}, 
\beqn
&&\frac{d\sigma^{qN(A)\to qgX}}{d^2\vec{k}_T} (k_T, x)= \frac{1}{(2\pi)^2}
\int d^2b~d^2r_1~d^2r_2 \nonumber\\
&\times&e^{i\vec k_T(\vec r_1-\vec r_2)}\overline{\Psi_{qg}^*(\vec r_1,\alpha )\Psi_{qg}(\vec r_2,\alpha )} \nonumber\\
&\times&
\Bigl[
\mathcal{N}^{N(A)}_{g\bar{q}q}(\vec b,\vec r_{1},\vec r_{1}-\alpha \vec r_2,x)
+
\mathcal{N}^{N(A)}_{g\bar{q}q}(\vec b,\vec r_{2},\vec r_{2}-\alpha \vec r_1,x)\nonumber\\
&-& \mathcal{N}^{N(A)}_{\bar{q}q}(\vec b,\alpha (\vec{r}_{1}-\vec{r}_{2}), x)-\mathcal{N}^{N(A)}_{gg}(\vec b,(\vec{r}_{1}-\vec{r}_{2}), x)\Bigr],\nonumber\\
\label{qg}
 \eeqn 
where $\vec r_{1}$ and $\vec r_{2}$ are the quark-gluon
 transverse separation in the direct and complex conjugated amplitude
 respectively. For brevity, we define again $\alpha$ as
 the fractional LC momentum of the radiated gluon, $\alpha=p_{+}(g)/p_{+}(q)\ll 1$.

In Eqs. (\ref{80},\ref{qg}) we have already integrated over the transverse coordinates of the second
parton. Note that the formulas in Eqs. (\ref{80},\ref{qg}) are given in impact parameter
representation and contain the sum of diagrams given in Fig.~\ref{f1}. The derivation of
these equations can be found in Refs. \cite{kst2,kst1}. The collinear
divergences which are the source of scale dependence of the parton
distribution functions and fragmentation functions in the
factorization Eq. (\ref{pp1}) are already subtracted in these equations.

The interaction amplitude of a colorless
 $g\bar{q}q$ and $gg$ system
 with a nucleon target can be written in
 terms of $\bar{q}q$ dipole amplitudes \cite{zakh},
\beqn
\mathcal{N}_{g\bar{q}q}^{N}(\vec b,\vec r_{1},\vec
r_{2},x)&=&\frac{9}{8}\Big\{\mathcal{N}^{N}_{\bar{q}q}(\vec
b,\vec{r}_{1},x)+\mathcal{N}^{N}_{\bar{q}q}(\vec
b,\vec{r}_{2},x)\Big\}\nonumber\\ &-& \frac{1}{8}
\mathcal{N}^{N}_{\bar{q}q}(\vec b,\vec{r}_{1}-\vec{r}_{2},x),\label{nqq0}\\ 
\mathcal{N}^{N}_{gg}(\vec
b,\vec r,x) &=&\frac{9}{4}\mathcal{N}^{N}_{\bar{q}q}(\vec
b,\vec{r},x). \label{nqq1}\
\eeqn
Again likewise Eq.~(\ref{3g-1}), the above equations immediately
satisfy several simple limiting cases. When the $q\bar{q}$ transverse
separation goes to zero i.e. $\vec r_1\approx \vec r_2$, the
$q\bar{q}$ pair is indistinguishable from a gluon, and
Eq.~(\ref{nqq0}) correctly reduces to
$\mathcal{N}_{g\bar{q}q}^{N}(\vec b,\vec r_{1},\vec
r_{1},x)=\frac{9}{4}\mathcal{N}^{N}_{\bar{q}q}(\vec
b,\vec{r}_{1},x)$. Moreover, in the limit of vanishing $\vec r_1$ (or
$\vec r_2$), the $qg$ (or $\bar{q}g$) system is indistinguishable from
a quark (antiquark) and Eq.~(\ref{nqq0}) becomes
$\mathcal{N}_{g\bar{q}q}^{N}(\vec b,\vec
r_{1},0,x)=\mathcal{N}^{N}_{\bar{q}q}(\vec b,\vec{r}_{1},x)$.

In the
derivation of Eqs.~(\ref{80}, \ref{qg}), one can rearrange the final
result in terms of Eqs.~(\ref{3g-1},\ref{nqq0},\ref{nqq1}).  However,
this is more than just some change of notation since the combination
of the right-hand sides of Eqs.~(\ref{3g-1},\ref{nqq0},\ref{nqq1}) are already well-known as
forward scattering amplitudes of $ggg$, $g\bar{q}q$ and $gg$ system
interacting with a proton target.  Eqs.~(\ref{3g-1},\ref{nqq0},\ref{nqq1}) are exact and are
not based on any approximation. A formal derivation of these equations
is similar to the derivation of the $q\bar{q}$-proton dipole
cross-section, namely one replaces the $q\bar{q}$ system by a
$q\bar{q}g$ (or $ggg$ , $gg$) one. The exchanged gluons can now couple
to different partons in the $q\bar{q}g$ system (or $ggg$ , $gg$) which
generates different phase factors. The precise calculation of the
color traces for the different couplings of the exchanged gluons to
the quark, antiquark and gluon ( or 3-gluons, gluon-gluon) leads to
the exact expression  given in Eqs.~(\ref{3g-1},\ref{nqq0},\ref{nqq1}) \cite{kst2,kst1,zakh,zakh1}.
Notice that in the CGC
approach the relations Eqs.~(\ref{3g-1},\ref{nqq0},\ref{nqq1}) holds only if one assumes that the
weight function for averaging over the target color charges are
Gaussian \cite{refr}.

The forward scattering amplitude of $\bar{q}q$, $g\bar{q}q$ and $gg$
interacting with a nucleus target at impact
parameter $\vec b$, can be again written, in eikonal form, in terms of
the dipole elastic amplitude $\mathcal{N}^{N}_{q\bar{q}}$ of a $\bar{q}q$ dipole
colliding with a proton at impact parameter $\vec{b}$,
\beqn
\mathcal{N}^{A}_{\bar{q}q}(\vec{b},\vec{r},x)&=&
1-e^{-\int d^{2}\vec{s}~
\mathcal{N}^{N}_{\bar{q}q}(\vec{s},\vec{r},x)T_{A}(\vec{b}+\vec{s})},\label{eik0}\\
\mathcal{N}^{A}_{g\bar{q}q }(\vec{b},\vec{r},x)&=&
1-e^{-\int d^{2}\vec{s}~
\mathcal{N}^{N}_{g\bar{q}q}(\vec{s},\vec{r},x)T_{A}(\vec{b}+\vec{s})},\label{eik00}\\
\mathcal{N}^{A}_{gg}(\vec{b},\vec{r},x)&=&
1-e^{-\frac{9}{4}\int d^{2}\vec{s}~
\mathcal{N}^{N}_{\bar{q}q}(\vec{s},\vec{r},x)T_{A}(\vec{b}+\vec{s})}\label{eik1}.\
\eeqn

The light-cone distribution of quark-gluon fluctuations $\Psi_{qg}$ in Eq.~(\ref{qg}) is given in Ref.~\cite{kst2}.
In the limit $\alpha\ll 1$ which is of practical interest at high energy, the quark-gluon distribution function including non-perturbative effects has the form,  
\begin{eqnarray}
\Psi_{qg}(\vec r, \alpha)=-\frac{2i}{\pi}\sqrt{\frac{\alpha_s}{3}}\frac{\vec{r}.\vec{e}^{\star}}{r^2}\exp\left(-r^{2}/2r_0^2\right). \label{w-qg}
\end{eqnarray}
where the parameter $r_0=0.3$ fm denotes the mean quark-gluon
separation and  is the result of a fit to soft diffraction $pp\to pX$.

One can show that for $\alpha\ll1$ the cross-section of gluon bremsstrahlung from
projectile quarks is $6$ times smaller than the corresponding
cross-section for a projectile gluon given by Eqs.~(\ref{80},\ref{100},\ref{qg},\ref{w-qg}) due to the 
color factor:
\begin{eqnarray}
\sigma^{qN(A)\to qgX}=\frac{\sigma^{gN(A)\to g_1g_2X}}{6}.
\end{eqnarray}

Note that similar results as Eqs.~(\ref{80},\ref{qg}) was also
obtained by Jalian-Marian and Kovchegov \cite{c1-n} in a color glass
condensate picture where the color dipole amplitudes in
Eqs.~(\ref{80},\ref{qg}) are replaced by a product of two Wilson lines
evaluated in the field of the color glass condensate. See also
Ref.~\cite{c2-n} for an earlier attempt along this line. Loosely
speaking, these two formulations are equivalent in the quasi-classical
(Glauber) approximation.  However, in order to include small-x
evolution, it is not sufficient to only put Wilson lines in the
evolved CGC fields. This only leads to logs of energy in the rapidity
interval between the produced gluon and the nucleus. One should also
include the evolution in the rapidity interval between the projectile
and the produced gluon \cite{c3-n}, thus describing gluon emission. It
was shown by Kovchegov and Tuchin \cite{c3-n} that such an evolution
is the linear BFKL equation due to some very interesting cancellations
of all nonlinearities.

In our approach, the effects of gluon emissions between the quark
(gluon) and the produced gluon (and its evolution) are effectively
included in the master Eqs.~(\ref{80},\ref{qg}) via the
non-perturbative quark-gluon (gluon-gluon) light-cone distribution
functions Eqs.~(\ref{100},\ref{w-qg}) which is obtained from a fit to
soft $pp$ diffraction data. The diffractive excitation of the incident
hadrons to the states of large mass is a more sensitive probe of
gluon-gluon fluctuations than the total cross section
\cite{kst2}. While the gluon emissions between the projectile and
target including their non-linear recombination effects are effectively
incorporated in terms of color-dipole forward amplitudes obtained from
a fit to DIS data.  By means of Eqs.~(\ref{80},\ref{qg}) one can also
describe the long-standing problem of the small size of the
triple-pomeron coupling \cite{kst2,spot}.

A word of caution is in order here. Notice that although the
non-perturbative $gg$ and $qg$ light-cone distribution functions include some saturation
effects of the projectile proton \cite{kst2,spot}. Nevertheless, the gluon
production cross-section given by Eqs.~(\ref{80},\ref{qg}) is
intrinsically asymmetric, namely it treats the "projectile" proton
approximately in a collinear factorization framework while treating
the "target" proton (or nucleus) in a saturation framework. 
Strictly speaking this may be justified only
in the case when saturation effects are present in the target wave
function, but are absent in the projectile wave function, such as in
$pA$ collisions or in forward particle productions. Although it
appears that such a simple approximation is sufficient to describe the
existing experimental data for hadron and direct photon production at
small $x$ at midrapidity in $pp$ collisions, see Figs.~\ref{rhic-pp},
\ref{phon}. Nevertheless, our formulation at midrapidity in $pp$
collisions is not well justified and therefore our results at
midrapidity in $pp$ collisions may not be valid.


\section{Hadrons production in high-energy $pp$ and $pA$ collisions}

The cross section of hadron production in $pp$ (or $pA$) collisions at impact
parameter $\vec b$ is given by a convolution of the distribution
function of the projectile gluon or quark inside the proton with the gluon
radiation cross-section coming from $gN$ or $qN$ ($gA$ or $qA$) collisions and also with the
fragmentation functions. 
For simplicity, we assume here that the projectile
gluon/quark has the same impact parameter relative to the target as the
beam proton. This is certainly a rather poor approximation which we will try to improve upon in future.

\begin{widetext}
\beqn
\frac{d \sigma^{pp(A)\to h+X}}{dy d^2 \vec p_T d^2 \vec b}&=& \int_{x_1}^{1} dz f_{g/p}(\frac{x_1}{z},Q^2) \frac{d\sigma^{gp(A)\to g_1g_2X}}{d^2k_{T}\,d^2b}(\frac{p_T}{z},\frac{x_2}{z}) \frac{D_{h/g_2}(z,Q^2)}{z^2} \nonumber\\
&+&\sum_{q,\bar{q}}\int_{x_1}^{1} dz f_{q/p}(\frac{x_1}{z},Q^2) \frac{d\sigma^{qp(A)\to qgX}}{d^2k_{gT}\,d^2b}(\frac{p_T}{z},\frac{x_2}{z})\frac{D_{h/q}(z,Q^2)}{z^2}\nonumber\\
&+&\sum_{q,\bar{q}}\int_{x_1}^{1} dz f_{q/p}(\frac{x_1}{z},Q^2) \frac{d\sigma^{qp(A)\to qgX}}{d^2k_{gT}\,d^2b}(\frac{p_T}{z},\frac{x_2}{z})\frac{D_{h/g}(z,Q^2)}{z^2},\nonumber\\
\label{pp1}
\eeqn
\end{widetext}
$f_{q/p}(x_q,Q^2)$ and $f_{g/p}(x_{g},Q^2)$ are the parton
distribution functions (PDF) of the colliding protons, which depend on the hard scale
$Q$ and the light-cone momentum fractions $x_{q}$ and $x_g$ for quarks and gluons, respectively.  The function $D_{h/q,g}(z,Q^2)$ is the fragmentation function
of parton $q,g$ to the final hadron $h$ with a momentum fraction
$z$. In the above equation, the variables $\frac{x_1}{z}$ and $\frac{x_2}{z}$ are momentum fractions of a parton in the beam and target.
The variables $x_{1,2}$ are defined by,  
\beq
x_1=\frac{p_T}{\sqrt{s}}e^{+\eta}, \hspace{2cm} x_2=\frac{p_T}{\sqrt{s}}e^{-\eta}, \label{x1x2}
\eeq
where $p_T$ and $\eta$ are the transverse momentum and rapidity of the produced hadron.

In Eq.~(\ref{pp1}) the cross-sections of gluon radiation in $gp(A)\to
ggX$ and $qp(A)\to qgX$ are given by
Eqs.~(\ref{80},\ref{qg}). We assume that the projectile parton
acquires high transverse momentum $k_T$ as a result of
coherent multiple rescattering, while the radiated gluons that
generate this momentum are summed to build up the color dipole
cross-section. Then, explicit inclusion of gluon bremsstrahlung balances the large $k_T$.

Notice that in the dipole approach in contrast to the parton model, one should
rely on the parton distribution functions taken at a soft scale since the evolution to the hard scale is performed via gluon radiation, which is
encoded in the phenomenological dipole cross-section fitted to DIS
data for the proton structure function. However, the dipole cross-section 
misses the $Q^2$-evolution of the $x_1$-distribution, which is especially important at forward
rapidities, since the parton distributions fall off at $x_1\to1$ much steeper at high $Q^2$. In
order to account for this effect and provide the correct
$x_1$-distribution, we take the integrated parton distribution in
Eq.~(\ref{pp1}) at the hard scale $Q=k_T$ \cite{amir00,amir1}.

Notice that at high energies and 
midrapidity the parton fractional momenta in the beam and target are
small, $x_1\sim x_2 \ll1$, so hadron production is dominated by
fragmentation of radiated gluons $gp(A)\to g_1g_2X$. However, at very
forward rapidities the quark contributions are important and the
subprocess $qp(A)\to qgX$ becomes relevant.  Therefore, different
subprocesses dominate in different kinematic regimes and their
overlap is small.


\section{Photon radiation in high-energy $pp$ and $pA$ collisions} 

Production of direct photons in the target rest frame should be
treated as electromagnetic bremsstrahlung by a quark interacting with
the target. In the light-cone dipole approach the transverse
momentum distribution of photon bremsstrahlung by a quark propagating
and interacting with a target nucleon (or nucleus $A$) at impact
parameter $b$, as calculated from the diagrams in Fig.~\ref{f1} (we show only 
the single gluon exchange diagrams), can be written in the factorized form  \cite{kst1,amir00,amir1}
\begin{eqnarray}
&&\frac{d \sigma(qN(A)\to\gamma X)}
{d(ln \alpha) d^{2}\vec{p}_T d^{2}\vec{b}}(\vec{p}_T,x)=\frac{1}{(2\pi)^{2}}
\sum_{in,f}\int d^{2}{r}_{1}d^{2}{r}_{2}\nonumber\\
&\times&
e^{i \vec{p}_T\cdot
(\vec{r}_{1}-\vec{r}_{2})}
\phi^{\star}_{\gamma q}(\alpha, \vec{r}_{1})
\phi_{\gamma q}(\alpha, \vec{r}_{2})~\Bigl[\mathcal{N}^{N(A)}_{q\bar{q}}(\vec{b},\alpha\vec{r}_{1},x) \nonumber\\
&+&\mathcal{N}^{N(A)}_{q\bar{q}}(\vec{b}, \alpha\vec{r}_{2},x)
-\mathcal{N}^{N(A)}_{q\bar{q}}(\vec{b},\alpha(\vec{r}_{1}-\vec{r}_{2}),x)\Bigr],\
 \label{m1}
\end{eqnarray} 
where $\vec{r}_{1}$ and $\vec{r}_{2}$ are the quark-photon transverse
separations in the direct and complex conjugated amplitudes
respectively; $\alpha=p_\gamma^+/p_q^+$ denotes the fractional
light-cone (LC) momentum of the radiated photon.  Correspondingly, the
transverse displacements of the recoil quarks in the two amplitudes
are $\alpha r_{1}$ and $\alpha r_{2}$ respectively. In Eq.~(\ref{m1}), $\phi_{\gamma q}(\alpha, \vec{r})$ is the
light-cone (LC) distribution amplitude of the projectile quark $\gamma
q$ fluctuation. Averaging over the initial quark polarizations and
summing over all final polarization states of the quark and photon, we get
\begin{eqnarray}
&&\sum_{in,f}\phi^{\star}_{\gamma q}(\alpha, \vec{r}_{1})\phi_{\gamma q}(\alpha, \vec{r}_{2})=\frac{\alpha_{em}}{2\pi^{2}}m^2_{q}\alpha^{2}
\Biggl\{\alpha^{2}K_{0}(\alpha m_{q} r_{1})\nonumber\\
&\times& K_{0}(\alpha m_{q} r_{2})+[1+(1-\alpha)^{2}]\frac{\vec{r}_{1}.\vec{r}_{2}}{r_{1}r_{2}}K_{1}( \alpha m_{q}r_{1})\nonumber\\
&\times&K_{1}(\alpha m_{q} r_{2}) \Biggl\} ,\label{wave}\
\end{eqnarray}
where $K_{0,1}(x)$ denotes modified Bessel functions of the second
kind and $m_{q}$ is an effective quark mass, which can be regarded as
a cutoff regularization.  Following Refs.~\cite{amir00,amir1,joerg} we take $m_{q}=0.2$ GeV. 
The forward scattering amplitude $\mathcal{N}^{A}_{q\bar{q}}$ can be
again written, in the eikonal form, in terms of the dipole
elastic amplitude $\mathcal{N}^{N}_{q\bar{q}}$ of a $\bar{q}q$ dipole
colliding with a proton at impact parameter $\vec{b}$ as defined in Eq.~(\ref{eik0}).

In order to obtain the hadron cross-section from the elementary
partonic cross section Eq.~(\ref{m1}), one should sum the
contributions from quarks and antiquarks (since only quarks and
antiquarks can radiate photons) weighted with the corresponding parton
distribution functions. The PDFs of the projectile enter in a
combination which can be written in terms of proton structure function
$F_{2}^{p}(x, Q^2)$. Notice that the contribution of gluon splitting
to quark-antiquark pairs (and higher Fock components) is already
contained in the sea quark distributions of the proton. Therefore, the
direct-photon production cross-section in $pp$ and $pA$ collisions is
given by \cite{joerg,amir00,amir1},
\begin{eqnarray}
&&\frac{d\sigma(pp(A)\to\gamma X)}{dx_{F}\, d^2\vec{p}_{T}\,d^2\vec{b}}=
\frac{x_{1}}{x_{1}+x_{2}}\int_{x_{1}}^{1}\frac{d\alpha}{\alpha^{2}}\nonumber\\
&\times&\sum Z_{f}^{2}\{q_{f}(\frac{x_{1}}{\alpha})+\bar{q}_{f}(\frac{x_{1}}{\alpha})\}
\frac{d \sigma(qp(A)\to\gamma X)}{d(ln \alpha) d^{2}\vec{p}_T d^{2}\vec{b}}(\vec{p}_T,x_2),\nonumber\\
&=&\frac{1}{x_1+x_2}
\int\limits_{x_1}^1 d\alpha \, F_{2}^{p}\left({x_1\over\alpha},Q^2\right)\, \frac{d \sigma(qp(A)\to\gamma X)}
{d(ln \alpha) d^{2}\vec{p}_T d^{2}\vec{b}}(\vec{p}_T,x_2),
 \label{con1}\nonumber\\
\end{eqnarray}
where the variable $x_1$ and $x_2$ are defined in Eq.~(\ref{x1x2}) and
$x_F=x_1-x_2$ is the Feynman variable. We have recently shown that in
this framework one can obtain a good description of the cross-section
for prompt photon production in proton-proton collisions at
RHIC and Tevatron energies \cite{amir00,amir1}, and Drell-Yan dilepton pair production \cite{amir00,jorg}. Here, we employ
this formulation to give predictions for the ratio of photon/pion production cross-sections at various
rapidities for LHC. We will also provide prediction for the nuclear modification factor in $pA$ collisions at LHC.

Notice that in the color-dipole factorization Eqs.~(\ref{pp1},\ref{con1}) neither K-factors (next-to-leading-order corrections), nor
higher twist corrections should be added. The phenomenological
dipole cross-section fitted to DIS data should already incorporate all perturbative
and non-perturbative radiation processes. The only contribution which is still missing in Eq. (\ref{con1}) is the
effect of the primordial momentum of the projectile parton. However,
it has been shown that in the color-dipole approach, the primordial
momentum should have a purely non-perturbative origin, and is
considerably smaller than in the parton model  \cite{amir00,amir1}. 
This effect should be of little importance for the kinematic regions 
of interest of this paper.

A word of caution is in order here. The type of factorization scheme
outlined above Eqs.~(\ref{pp1},\ref{con1}) has not been yet rigorously proven at any order of pQCD in the kinematic region of our
interest and is most probable not exact. Nevertheless, there is growing evidence in the literature that 
it gives a good approximation  for the processes discussed here \cite{break,kst2,kst1,boris1,me-rev,dhj,amir00,amir1,joerg,jorg}.


\section{ Gluon saturation and color dipole models}

At high energies/small Bjorken-x, QCD predicts that gluons in a hadron wavefunction form a new state, the so-called
Color Glass Condensate (CGC)
\cite{Gribov:1984tu1,Gribov:1984tu,Jalilian-Marian:1997jx,Kovchegov:1999yj,Iancu:2003xm}.
The cornerstone of the CGC is the existence of a hard saturation
scale $Q_s$ at which nonlinear gluon recombination effects become important and start to
balance gluon radiation. 

\begin{figure}[!t]
       \includegraphics[width=7 cm] {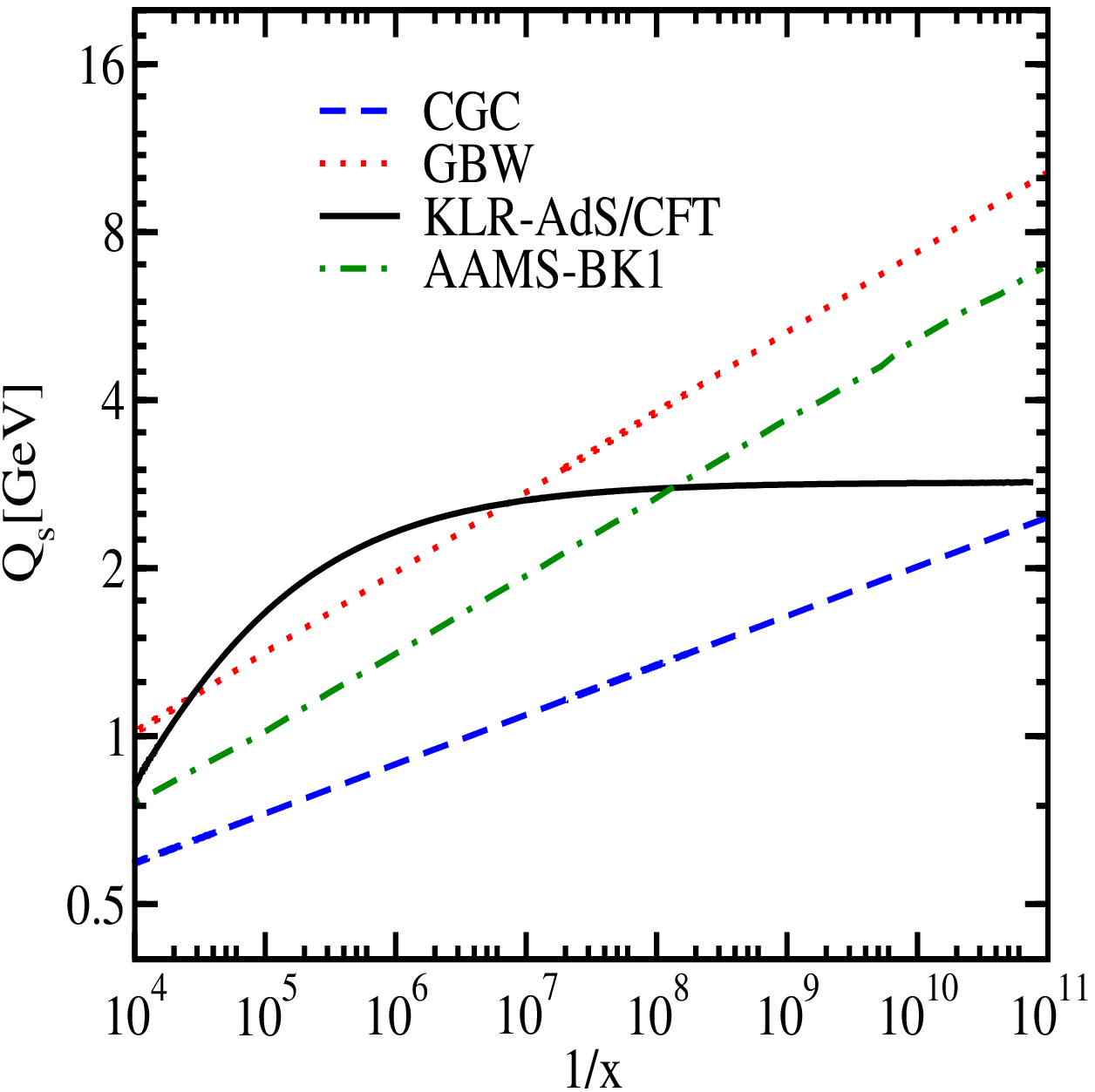}
       \includegraphics[width=7 cm] {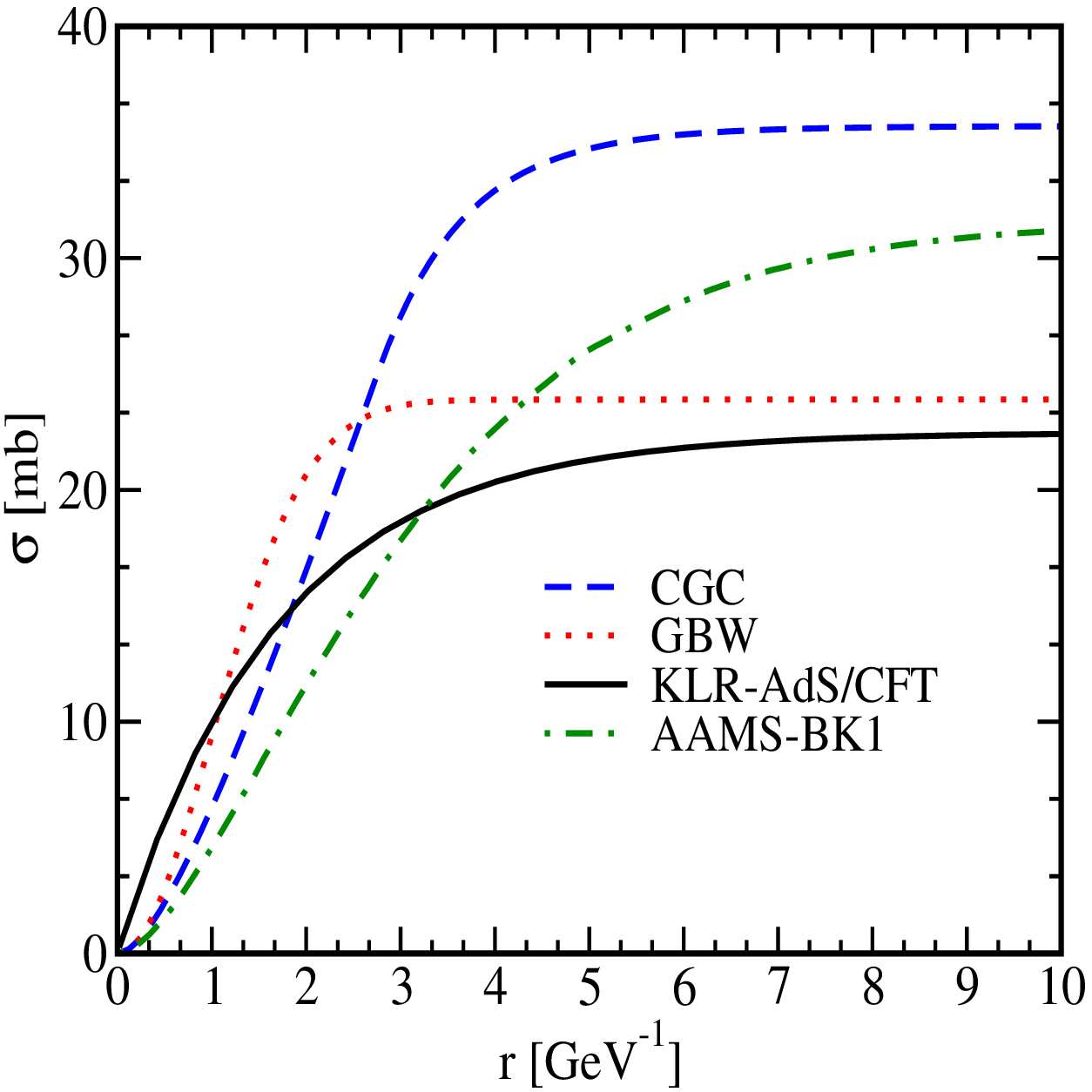} \caption{Top
       panel: Saturation scale defined via Eq.~(\ref{defs}) as a function of $1/x$ for various
       color-dipole models. Lower panel: The total dipole-proton cross
       section $ \sigma_{q\bar{q}}(r,x)$ at fixed $x=10^{-5}$ in the various
       color-dipole models introduced in Sec. VI. } \label{f:sat}
\end{figure}

The concept of saturation and the taming of the power-like rise of the
gluon distribution at small $x$ was first addressed by Gribov, Levin and Ryskin in
the double logarithmic approximation \cite{Gribov:1984tu1}. A first hint toward saturation
effects at HERA came from the phenomenologically success of the
Golec-Biernat and W\"usthoff (GBW) model \cite{gbw}. This model
incorporates the basic saturation effects into the
color-dipole cross-section on a proton target.  In the CGC framework
the dipole-proton forward scattering amplitude can be in principle
found by solving the perturbative nonlinear small-x Balitsky-Kovchegov
(BK) \cite{Kovchegov:1999yj} or
Jalilian-Marian--Iancu--McLerran--Weigert--Leonidov--Kovner (JIMWLK)
\cite{Jalilian-Marian:1997jx} quantum evolution equations.  The BK and
JIMWLK evolution equations unitarize the linear
Balitsky-Fadin-Kuraev-Lipatov (BFKL) \cite{Kuraev:1977fs} evolution
equation at small-$x$ in the large-$N_c$ limit (BK) and beyond (JIMWLK).
It has been shown that next-to-leading-order
(NLO) corrections to the BFKL equation (and therefore to BK and JIMWLK
kernels) are large and negative \cite{Fadin:1998py}. 
There was no reason to believe that still higher order corrections are
unimportant, until quite recently, when it was found that the
consistent incorporation of the running coupling $\alpha_s$ into the BFKL, BK and JIMWLK equations
\cite{Albacete:2007sm,rc,Albacete:2009fh} leads to phenomenologically rather successful descriptions.   
Still the actual calculation of higher-order corrections to 
these non-linear evolution equations remains as a challenge. 
Thus, we resort to a QCD-like model which incorporates the basic
features of gluon saturation into the dipole-proton forward scattering
amplitude, and provides predictions which will allow to test the validity of our treatment. There are several parametrizations
proposed in the literature which all give a good description of HERA
data but predict different saturation scales, see Fig.~\ref{f:sat}. In this section we review some
of these models and later we will employ them for hadron and photon production
in various kinematic regimes and investigate the uncertainties of the various models and discuss the 
differences between them.

\subsection{GBW model}
The dipole-proton cross-section $\sigma_{q\bar{q}}(r,x)$ is usually written as an integral of the
imaginary part of the forward scattering amplitude $\mathcal{N}_{q\bar{q}}^{N} (\vec{r}, \vec{b}, s)$ over the impact parameter $\vec{b}$ as defined via Eq.~(\ref{di-app}). 
One may neglect the $\vec b$-dependence in $\mathcal{N}_{q\bar{q}}^{N}$ making the integral in Eq.~(\ref{di-app}) trivial, giving the proton's
transverse area factor: 
\begin{equation}
\sigma_{q\bar{q}}(r,x) \equiv \sigma_0 \, \mathcal{N}_{q\bar{q}}^{N}(r,x). \label{s-dd}
\end{equation}
A popular parametrization for the $q\bar{q}$ dipole cross-section on a nucleon target is due to Golec-Biernat and W\"usthoff (GBW)
\cite{gbw} and is able to describe
DIS data with a simple form for the color dipole amplitude, 
\begin{equation}
\mathcal{N}_{q\bar{q}}^{\text{GBW}}(r,x)=1-e^{-r^{2}Q_{s}^{2}(x)/4}, \label{gbw1}
\end{equation}
where the $x$-dependence of the saturation scale is given by 
\begin{equation}
Q_s^2(x)=(x_0/x)^{\lambda}~\text{GeV}^2. \label{sat-s}
\end{equation}
The main feature of the model is that for decreasing $x$, the dipole
amplitude saturates at smaller dipole sizes.  Note that there is no
unique definition for the saturation scale in literature. Following
Refs.~\cite{gbw,di1,di2,di3} we define the saturation scale
$Q_s^2=2/r_s^2$ as a energy scale at which the $q\bar{q}$ dipole
scattering amplitude $\mathcal{N}$ becomes sizable,
\begin{equation}
\mathcal{N}_{q\bar{q}}(r_s=\sqrt{2}/Q_s,x)\equiv 1 - e^{-1/2} \approx 0.4.  \label{defs}
\end{equation}
For the GBW model, this definition coincides with the saturation scale
$Q_s$ defined in Eq.~(\ref{sat-s}). The value of the intercept
$\lambda\approx 0.25-0.30$ is consistent with perturbative predictions
based on small-x evolution \cite{lqcd,Albacete:2007sm,Albacete:2009fh,Iancu:2002tr,lbk}. The
parameters $\sigma_{0}=23.9$ mb, $x_{0}=1.11\times 10^{-4}$, and
$\lambda=0.287$ were determined from a fit to $F_2$ for $x<0.01$
and $Q^2\in[0.25,45]$ in the presence of charm quarks with mass
$m_c=1.4$ GeV \cite{di2}. Note that the saturation scale in the GBW model reduces with
the inclusion of the charm quark \cite{di2}.

\subsection{CGC, AAMS-BK and b-CGC models}
The linear DGLAP evolution equation which only includes gluon radiation may not
be appropriate for the saturation regime where nonlinear recombination
subprocess are important.  Iancu, Itakura and Munier proposed an
alternative color glass condensate (CGC) model \cite{CGC0}, based on
the BK equation \cite{Kovchegov:1999yj}. In this model the $q\bar{q}$ dipole amplitude for a nucleon target is parametrized as,
\begin{equation} \label{cgc}
  \mathcal{N}_{q\bar{q}}^{\text{CGC}}(r,x) =
  \begin{cases}
    \mathcal{N}_0\left(\frac{rQ_s}{2}\right)^{2\left(\gamma_s+
    \frac{1}{\kappa\lambda Y}\ln\frac{2}{rQ_s}\right)} & :\quad rQ_s\le 2\\
    1-\mathrm{e}^{-A\ln^2(BrQ_s)} & :\quad rQ_s>2
  \end{cases},
\end{equation}
where the saturation scale is again parametrized as Eq.~(\ref{sat-s}), $Y=\ln(1/x)$, and $\kappa = \chi''(\gamma_s)/\chi'(\gamma_s)$ where $\chi$ is the LO BFKL characteristic function.  
The coefficients $A$ and $B$ in the second line of \eqref{cgc} are determined uniquely from the condition that the color dipole cross-section 
and its derivative with respect to $rQ_s$ are continuous at $rQ_s=2$:
\begin{equation} \label{eq:AandB}
  A = -\frac{\mathcal{N}_0^2\gamma_s^2}{(1-\mathcal{N}_0)^2\ln(1-\mathcal{N}_0)}, \qquad B = \frac{1}{2}\left(1-\mathcal{N}_0\right)^{-\frac{(1-\mathcal{N}_0)}{\mathcal{N}_0\gamma_s}}.
\end{equation}
The parameters $\gamma_s=0.63$ and $\kappa=9.9$ are fixed at the LO
BFKL values. The others parameters $\mathcal{N}_0=0.7$,
$\sigma_0=35.7$ mb, $x_0=2.7\times 10^{-7}$ and $\lambda=0.177$ were
fitted to $F_2$ for $x<0.01$ and $Q^2<45$ $\text{GeV}^2$ and including
a charm quark with $m_c=1.4$ GeV. Notice that for small $rQ_s\le 2$,
the effective anomalous dimension $1-\gamma_s$ in the exponent in the
upper line of Eq.~(\ref{cgc}) rises from the LO BFKL value towards the
DGLAP value. 

Recently, Albacete, Armesto, Milhano and Salgado calculated
numerically the dipole-proton scattering amplitude from the BK
equation including running coupling corrections (AAMS-BK1,2 model) \cite{Albacete:2009fh}. Note
that the incorporating of the running coupling is essential in this
approach, though its implementation is model dependent. The free
parameters in their fit to HERA data are related to the initial
condition for the evolution at $x_{in}=10^{-2}$. They used two families
of initial conditions, the GBW form (AAMS-BK1 model)
\begin{equation}
 \mathcal{N}_{in}^{GBW}(r,x_{in})=
1-\exp{\left[-\left(\frac{r^2\,Q_{s\,0}^2}{4}\right)^{\gamma\,}\right]}\,,
\label{gbw-i}
\end{equation} 
and the McLerran-Venugopalan form (AAMS-BK2 model):
\begin{equation} 
\mathcal{N}_{in}^{MV}(r,x_{in})=1-\exp{\left[-\left(\frac{r^2Q_{s\,0}^{2}}{4}\right)^{\gamma}
    \ln{\left(\frac{1}{r\,\Lambda_{QCD}}+e\right)}\right]}\, ,
\label{mv}
\end{equation} 
where $Q_{s\,0}^2$ is the initial saturation scale. 
In their global analysis of HERA data there are four free
parameters which are fitted to $F_2$-data for $x\le 0.01$ and $Q^2/\text{GeV}^2\in [0.045,800]$: the initial saturation scale $Q_{s0}$, the overall
normalization $\sigma_0$, the infrared parameter $C$ introduced in the running coupling and the anomalous
dimension $\gamma$. The values of parameters can be found in table $1$ of Ref.~\cite{Albacete:2009fh}.

The gluon density is larger in the center of a proton $b=0$ than at
periphery $b\sim 2-3~\text{GeV}^{-1}$ probed in the total
$\gamma^{\star}p$ cross-section. Therefore, impact-parameter
dependence of the dipole-proton forward scattering amplitude seems to be essential. There has been several attempts to
model the impact-parameter dependence in dipole-proton forward
scattering amplitudes. We consider here the model proposed by Watt and
Kowalski (b-CGC) \cite{bccc}. In this model, the dipole-proton forward
scattering amplitude has the same form as the CGC model Eq.~(\ref{cgc}), but 
the saturation scale $Q_s$ now depends on impact parameter,
\begin{equation} \label{bcgc1}
  Q_s\equiv Q_s(x,b)=\left(\frac{x_0}{x}\right)^{\frac{\lambda}{2}}\;\left[\exp\left(-\frac{b^2}{2B_{\rm CGC}}\right)\right]^{\frac{1}{2\gamma_s}}. 
\end{equation}
The parameter $B_{\rm CG}=7.5 \text{GeV}^{-2}$ is fitted to the
$t$-dependence of exclusive $J/\Psi$ photoproduction. It has been
shown that if one allows the parameter $\gamma_s$ to vary together 
with the other parameters (in contrast to the CGC fitting procedure where
$\gamma_s$ is fixed to its LO BFKL value), this results in a significantly
better description of data for  $F_2$ with the value of
$\gamma_s=0.46$, which is remarkably close to the
value of $\gamma_s=0.44$ recently obtained from the BK equation \cite{bkk}.
Other parameters obtained from the fit are: $\mathcal N_0=0.558$,
$x_0=1.84\times 10^{-6}$ and $\lambda=0.119$ \cite{bccc}.

Notice that calculation of the $p_T$-distribution
of produced hadrons/photons in $pp$ collisions needs only knowledge of
the total dipole cross-section and is independent of the
impact-parameter dependence of the forward scattering  dipole-proton
amplitude. Nevertheless, the integrated dipole cross-section of the b-CGC model is different from other dipole models


\subsection{KLR-AdS/CFT model}

The above mentioned dipoles models are motivated by pQCD and their
validity at very small $Q^2$ where one has to consider
small-$x$ evolution in the large coupling limit is questionable. Performing
calculations in the strong coupling limit of QCD is very difficult. One may
resort to other QCD-like theories, such as $\mathcal{N}=4$ Super-Yang-Mills where one can perform
calculations in the non-perturbative limit of large `t Hooft coupling
by employing the Anti-de Sitter space/conformal field theory (AdS/CFT)
correspondence \cite{Maldacena:1997re}. On this line, recently, Kovchegov, Lu and
Rezaeian \cite{amir-ads} proposed a new color dipole parametrization inspired by
the AdS/CFT approach (KLR-AdS/CFT) which reasonably well describes the HERA data for inclusive structure functions at small-$x$ and $Q^2$. In this model, the dipole-proton scattering amplitude is given by, 
\begin{widetext}
\begin{eqnarray}
  \mathcal{N}^{AdS}_{q\bar{q}}(r,x)& =&1-\exp\Big[-\frac{\mathcal{A}_0 \, x \, r}{\mathcal{M}_0^2(1-x)\pi \sqrt{2}}\left(\frac{1}{\rho_m^3}+\frac{2}{\rho_m}-2\mathcal{M}_0\sqrt{\frac{1-x}{x}}\right)\Big], \label{fas1}\
\end{eqnarray}
\end{widetext}
with notations
\begin{eqnarray}
\rho_m&=&  \begin{cases}
 (\frac{1}{3m})^{1/4}\sqrt{2\cos(\frac{\theta}{3})}  & : m \le \frac{4}{27} \nonumber\\
     \sqrt{\frac{1}{3m\Delta} +  \Delta} & :m> \frac{4}{27}
  \end{cases},  \nonumber\\
\Delta&=& \Big[\frac{1}{2m}-\sqrt{\frac{1}{4m^2}-\frac{1}{27m^3}}\Big]^{1/3}, \nonumber\\
m&=&\frac{\mathcal{M}_0^4(1-x)^2}{x^2}, \nonumber\\
\cos(\theta)&=&\sqrt{\frac{27m}{4}}. \label{not}\
\end{eqnarray}
where $\mathcal{A}_0=\sqrt{\lambda_{YM}} \, \text{GeV}$. The
parameters of the model for quark mass $m_q=140~\text{MeV}$ 
and `t Hooft coupling $\lambda_{YM}=10$  obtained
from the fit to the HERA data (in the range of $x
\in[6.2\times 10^{-7},6\times 10^{-5}]$ and $Q^2/\text{GeV}^2\in [0.045,2.5]$) are: 
$\mathcal{M}_0=8.16\times 10^{-3}$ and $\sigma_0=26.08~\text{mb}$ (see Eq.~(\ref{s-dd})). We
will also consider another fit to the same data but with `t Hooft
coupling $\lambda_{YM}=20$ which also gives a good
fit: $\mathcal{M}_0= 6.54 \times 10^{-3}$ and $\sigma_0=22.47~\text{mb}$ \cite{amir-ads}.

Similarly, the saturation scale in the KLR-AdS/CFT dipole model
(\ref{fas1}) can be obtained from the definition given in Eq.~(\ref{defs}),
\begin{equation} 
Q_s^{\text{AdS}}(x)=\frac{2 \, \mathcal{A}_0 \, x}{\mathcal{M}_0^2 \, (1-x) \, \pi} \, 
\left(\frac{1}{\rho_m^3}+\frac{2}{\rho_m}-2\mathcal{M}_0\sqrt{\frac{1-x}{x}}\right). \label{qads}
\end{equation}
In this model the saturation scale varies in the range of $1 \div
3~\text{GeV}$ becoming independent of energy/Bjorken-$x$ at very small
$x$ (see Fig.~\ref{f:sat}). This leads to the prediction of $x$-independence of the $F_2$
structure function at very small $x$ and $Q^2$ in a region where there is no experimental data yet. 

Note that the KLR-AdS/CFT dipole scattering amplitude exhibits the
property of geometric scaling \cite{Stasto:2000er}: it is a function
of $r \, Q_s^{\text{AdS}}(x)$ only.  Moreover, the anomalous dimension
in this model is $\gamma_s=0.5$ which is rather close to the value of
$0.44$ obtained from the numerical solution of the BK equation
\cite{bkk}. Thus in many ways the predictions of the KLR-AdS/CFT model are similar to the
predictions of the CGC model. Therefore, the non-perturbative
KLR-AdS/CFT model which is valid at low $Q^2<2.5~\text{GeV}^2$ could
be viewed as complementary to the perturbative description of data
based on saturation/Color Glass Condensate physics. The main
difference is the $x$-dependence of the saturation scale
$Q_s^{\text{AdS}}(x)$, which leads to $x$-scaling at small
$x$ and $Q^2$.

\subsection{Semi-Sat Model}
 In order to demonstrate the importance of saturation, we will
 also use a semi-saturation model (Semi-Sat) fitted to $F_2$ with $x\le 0.01$ and 
$Q^2 \in[0.25, 45]~ \text{GeV}^2$: 
 \begin{equation} 
\mathcal{N}_{q\bar{q}}^{\text{Semi-Sat}} (\vec{r}, \vec{b}, x) =2\mathcal{N}_0\left(\frac{rQ_s}{2}\right)^{2\gamma_{eff}},  \label{nos}
\end{equation}
where $Q_s$ is defined in Eq.~(\ref{bcgc1}). The parameter $\gamma_{eff}$ is defined for $rQ_s\le 2$ as $\gamma_{eff}=\gamma_s +\frac{1}{\kappa\lambda Y}\ln\frac{2}{rQ_s}$,
and for $rQ_s> 2$ as $\gamma_{eff}=\gamma_s$. The other parameters are given by
$\gamma_s=0.43$, $\mathcal N_0=0.568$, $x_0=1.34\times 10^{-6}$ and
$\lambda=0.109$ \cite{bccc}.  Surprisingly, the fit obtained with such
an oversimplified model is as good as for the other models with
$\chi^2/\text{d.o.f.}=0.92$.

Comparing \eq{cgc} and \eq{nos} one can see that they treat the
region $r\,Q_s \,>\,1$ differently. The CGC model describes this
region based on solutions to the BK equation \cite{MUT,IIM,LT} for
$r\,Q_s >\,2$ (with a phenomenological matching at $r\,Q_s = 2$) which
are also applied (somewhat inconsistent) in this model for $r$ close
to $1/Q_s$.

In Fig.~\ref{f:sat}, we show the saturation scale (top panel) and
$q\bar{q}$ dipole-proton cross-section (lower panel) within various
color-dipole models fitted to the HERA data. Note that we used for all
curves in Fig.~\ref{f:sat} the same definition for the saturation
scale given in Eq.~(\ref{defs}).  It is obvious that the discrepancies among
different models fitted to the same data are quite significant.
Therefore, it seems that HERA data alone is not sufficient for a satisfactory understanding of saturation physics. One of the aims of this
paper is to investigate if hadrons and photon production at LHC can improve our understanding of
saturation effects.

 \begin{figure}[!t]
       \centerline{\includegraphics[width=8 cm] {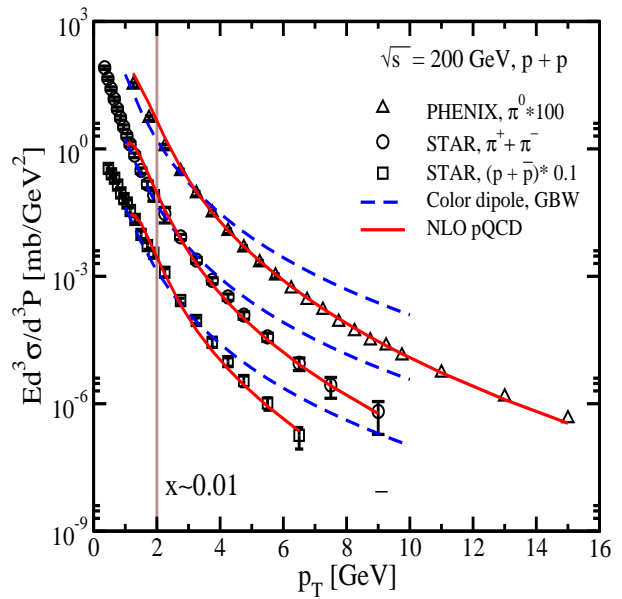}}
       \caption{ The hadrons spectra at the RHIC energy at
       midrapidity. The dashed lines are calculated with the GBW
       model. The solid lines are the pQCD calculation results taken
       from Ref.~\cite{me-croin}. Note that the color dipole approach is valid at very
       small $x_2$ corresponding to $p_T<2$ GeV at RHIC energy
       and midrapidity(shown by a line). The experimental data are from \cite{rhic2006,star-data}. \label{rhic-pp} }
\end{figure}


\section{Numerical results for $pp$ collisions}

In order to analytically reduce the four-dimensional integrals in the
partonic cross-sections Eqs.~(\ref{80},\ref{qg}) to one-dimensional
integrals Eqs.~(\ref{vn1},\ref{ga}), we assumed that the strong coupling $\alpha_s$ is a
constant. In principle, the strong coupling $\alpha_s$ entering in 
the $gg$ and $gq$ light-cone distribution functions of the incoming parton defined in 
Eqs.~(\ref{100},\ref{w-qg}) is a function of the transverse dipole size. To improve
our description, we replace $\alpha_s$ by $\alpha_s(k_T)$, where $k_T$ is the transverse momentum of the parton.
More precisely, in Eqs.~(\ref{80},\ref{qg}) we replace $\alpha_s(r_1)\alpha_s(r_2)\to \alpha_s^2(k_T)$ where 
$r_1$ and $r_2$ are the gluon-gluon (or quark-gluon) transverse
separation in the direct and complex conjugated amplitudes
respectively and are related by a double Fourier transformation
 to the transverse momentum of the radiated gluon $k_T$, see Eqs.~(\ref{80},\ref{qg}).

We employ recent NLO parton distribution functions (PDFs) developed
for LHC application (MSTW2008) \cite{mstw}. For the fragmentation functions
(FFs) we use the result of a recent NLO AKK08 analysis \cite{akk}.  For the running strong coupling
$\alpha_s$, we employ the same scheme as used for the MSTW2008 PDFs,
namely we solve the renormalization group equation in the MSbar scheme
at NLO level \cite{alphas}.  We stress that all phenomenological parameters in
our model are already fixed by other reactions and in this sense our
results can be considered as parameter-free predictions. 
\begin{figure}[!t]
       \includegraphics[width=8 cm] {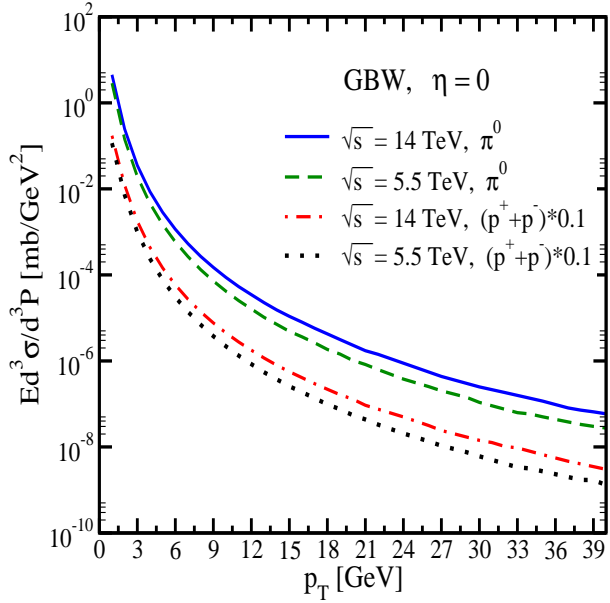} \caption{ Pions
       and protons spectra in $pp$ collisions at the LHC energies
       $\sqrt{s}=5.5$ and $14$ TeV at midrapidity. Theory curves are
       calculated with the GBW model. Note that our model for particle productions is not reliable at
midrapidity, for explanation see the end of Sec. III.\label{flhc-1}}
\end{figure}
\begin{figure}[!t]
       \includegraphics[width=6 cm] {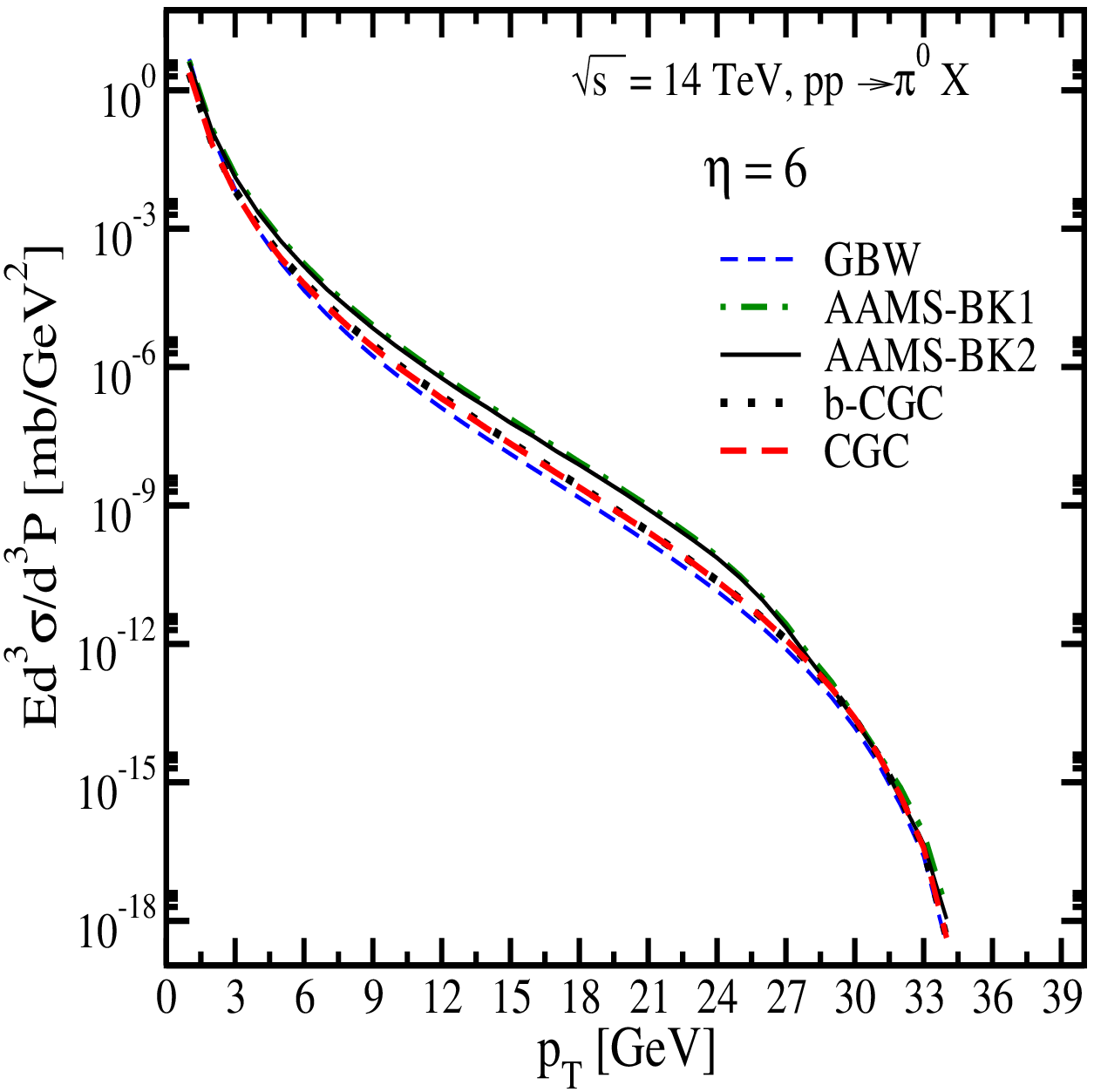}
       \includegraphics[width=6 cm] {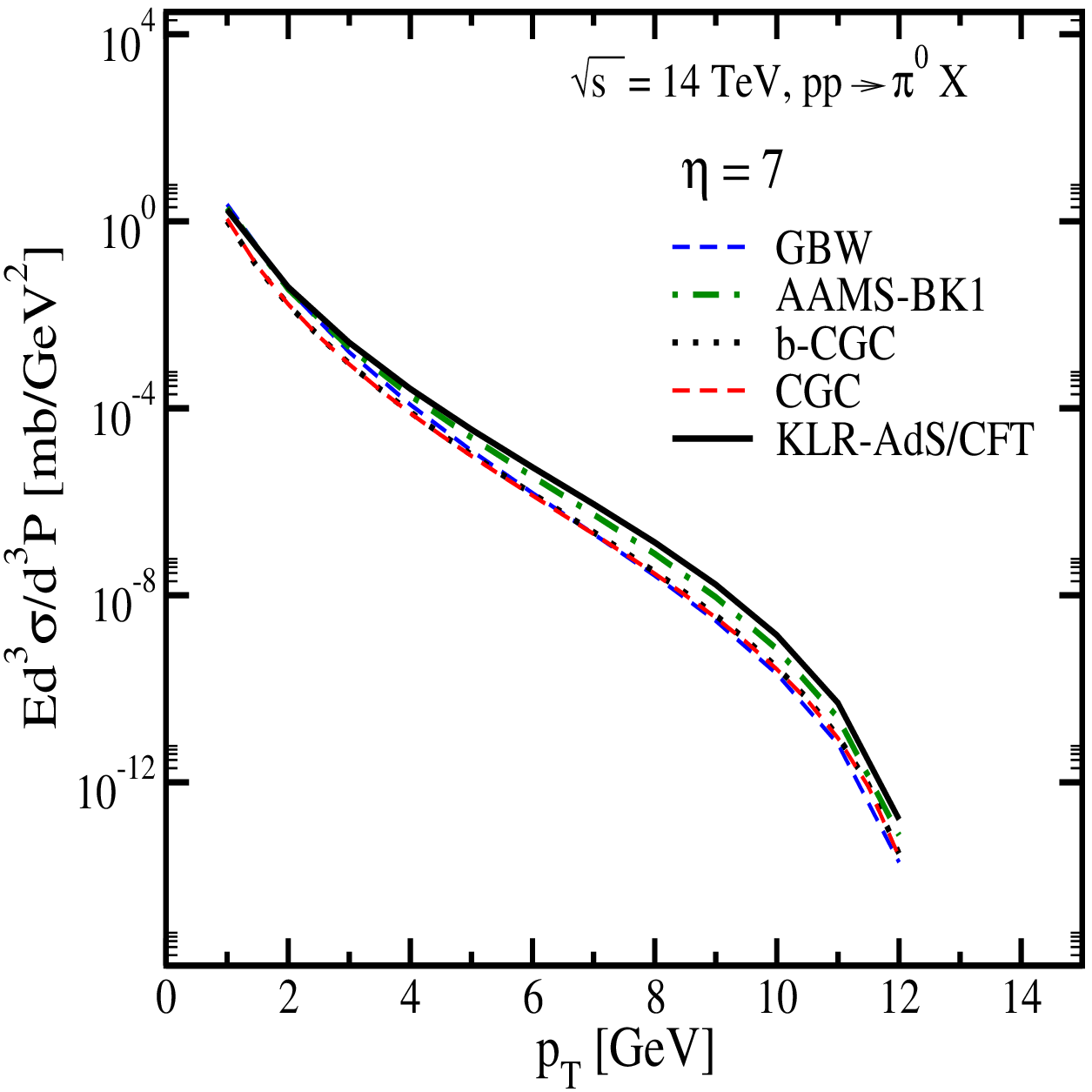}
        \includegraphics[width=6 cm] {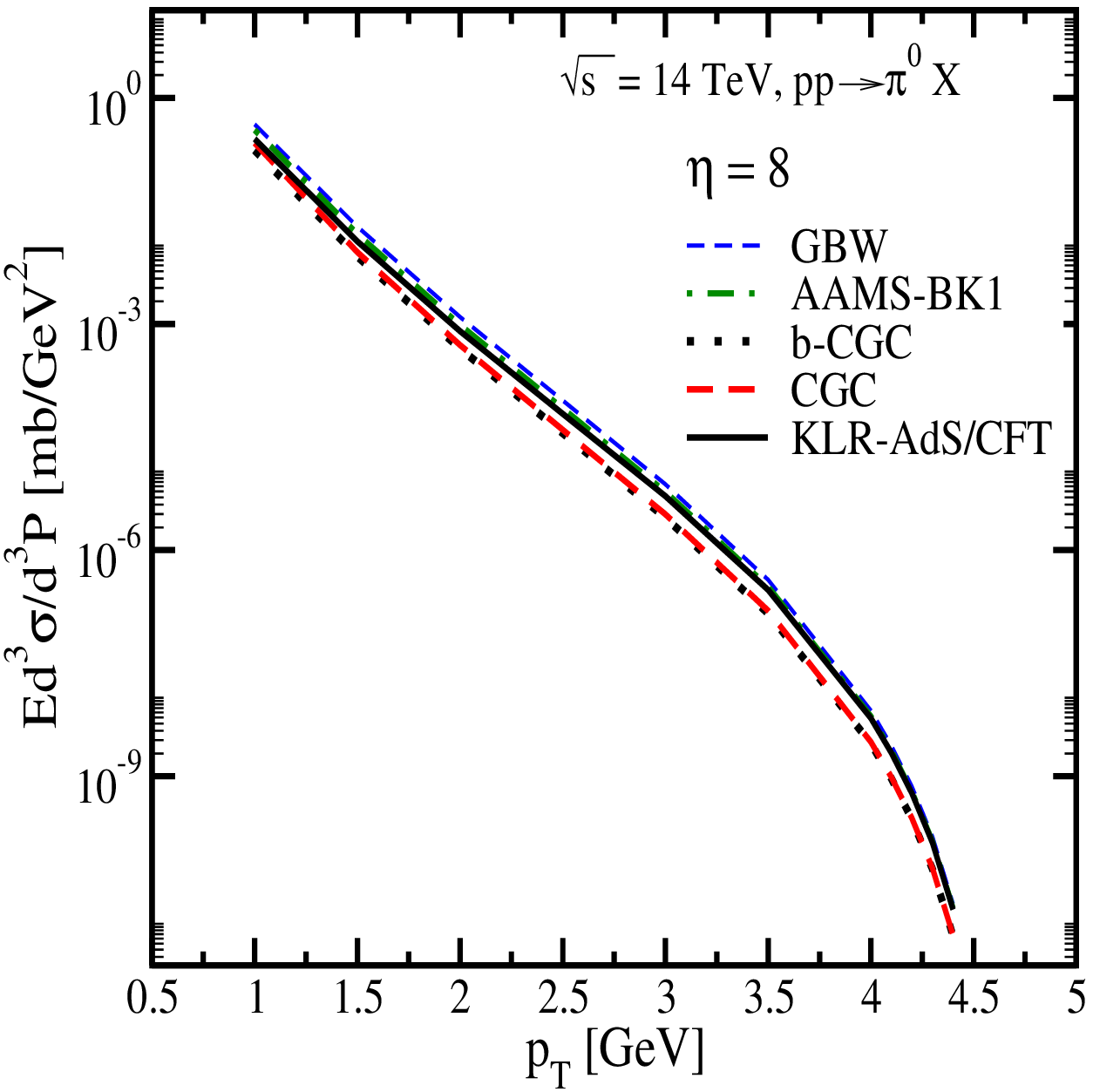}
       \caption{ Pion spectra obtained from various dipole models at forward rapidities and LHC energies for $pp$ collisions. \label{flhc-eta}}
\end{figure}

In Fig.~\ref{rhic-pp}, we show dipole model results obtained from the
light-cone factorization in Eq.~(\ref{pp1}) for pion ($\pi^0,
\pi^++\pi^-)$ and proton ($p+\bar{p}$) spectra at RHIC energy
$\sqrt{s}=200$ GeV and midrapidity. The experimental data are from
PHENIX \cite{rhic2006} and STAR \cite{star-data}. For a compassion, we also show the results coming
from an improved pQCD calculation performed in Ref.~\cite{me-croin}. Notice that in the parton model
results shown in Fig.~\ref{rhic-pp} a fixed $K$-factor $K=1.5$ was
introduced in order to simulate higher order perturbative corrections while in the
color dipole approach we do not introduce a $K$-factor since the
dipole-proton cross-section fitted to HERA incorporates all higher
order radiations.  Note that all the above-mentioned parametrizations for the color dipole
cross-section have been fitted to DIS data at $x\le 0.01$. This
corresponds to $p_T \le 2$ GeV for RHIC energy at midrapidity (see Eq.~(\ref{g-2g})), so
the PHENIX and STAR data plotted in Fig.~\ref{rhic-pp} are not suited
for a model test. It is seen from Fig.~\ref{rhic-pp} that deviation of
color dipole results from the experimental data starts at about $p_T=2-4$
GeV. At LHC energies $\sqrt{s}=5.5$ and $14$ TeV for a large range of $p_T$ (even at $\eta=0$) we have $x_2\ll 0.01$, therefore we expect the color dipole
prescription to be valid. In Fig.~\ref{flhc-1}, we show the predictions
of the GBW model for pion spectra in $pp$ collisions for LHC energies $\sqrt{s}=5.5, 14
$ TeV at midrapidity $\eta=0$. The predictions for pion invariant cross-sections at various rapidities in $pp$ collisions for LHC are given in 
Fig.~\ref{flhc-eta}. One can see from Fig.~\ref{flhc-eta} that various dipole models presented in
the previous section with explicit saturation give rather similar
results (we will scrutinize this below). Note that the KLR-AdS/CFT model described in Sec. VI-C was fitted to the HERA data with $x
\in[6.2\times 10^{-7},6\times 10^{-5}]$ and $Q^2/\text{GeV}^2\in [0.045,2.5]$. Therefore it is only valid at very forward rapidities and low $p_T$. 
As it is seen in the upper panel of Fig.~\ref{flhc-eta}, the two color dipole solutions of
the BK equation for the GBW and MV initial conditions (AAMS-BK1,2)
give very similar results and further on we will only
consider one of them.

\begin{figure}[!t]
       \includegraphics[width=6 cm] {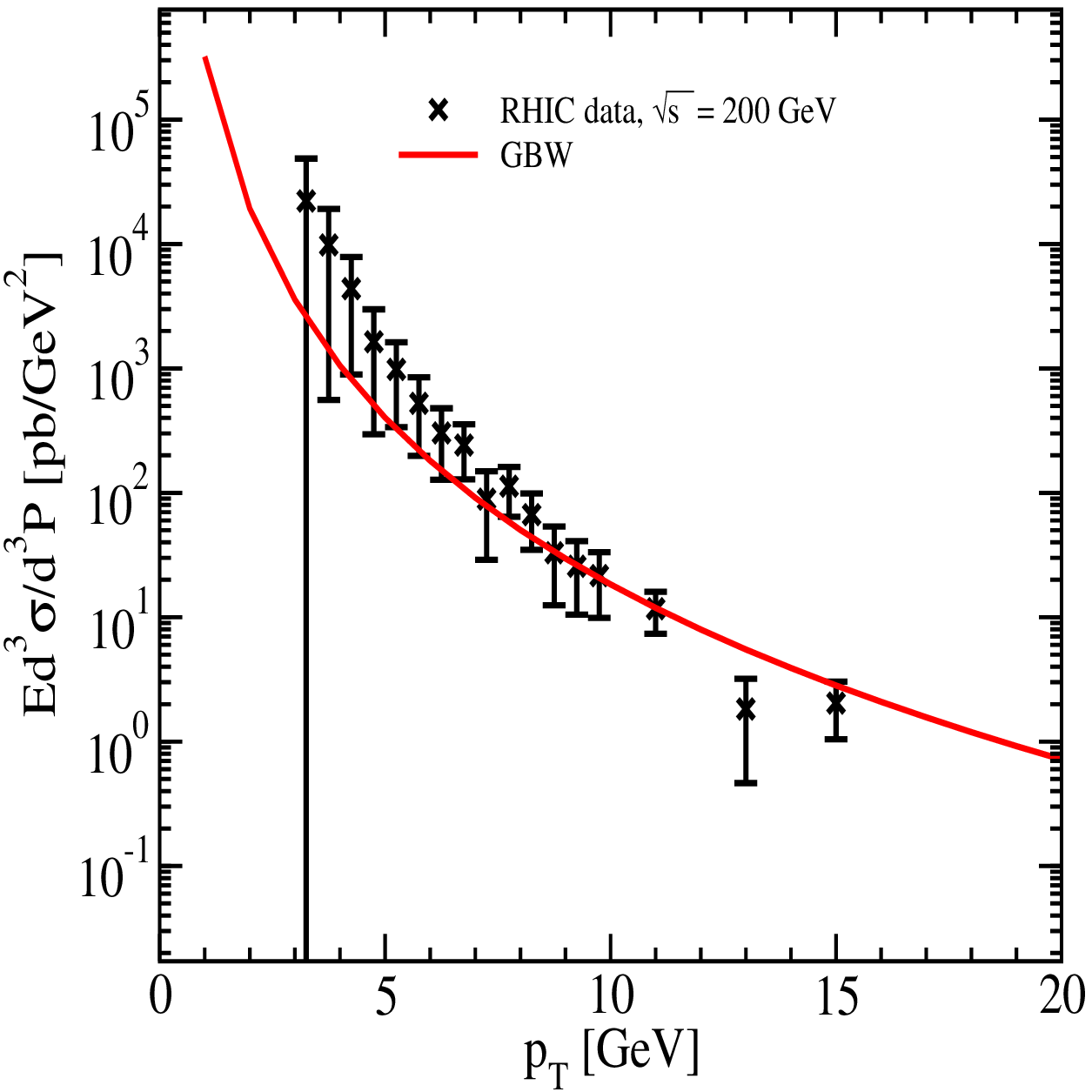}
        \includegraphics[width=6 cm] {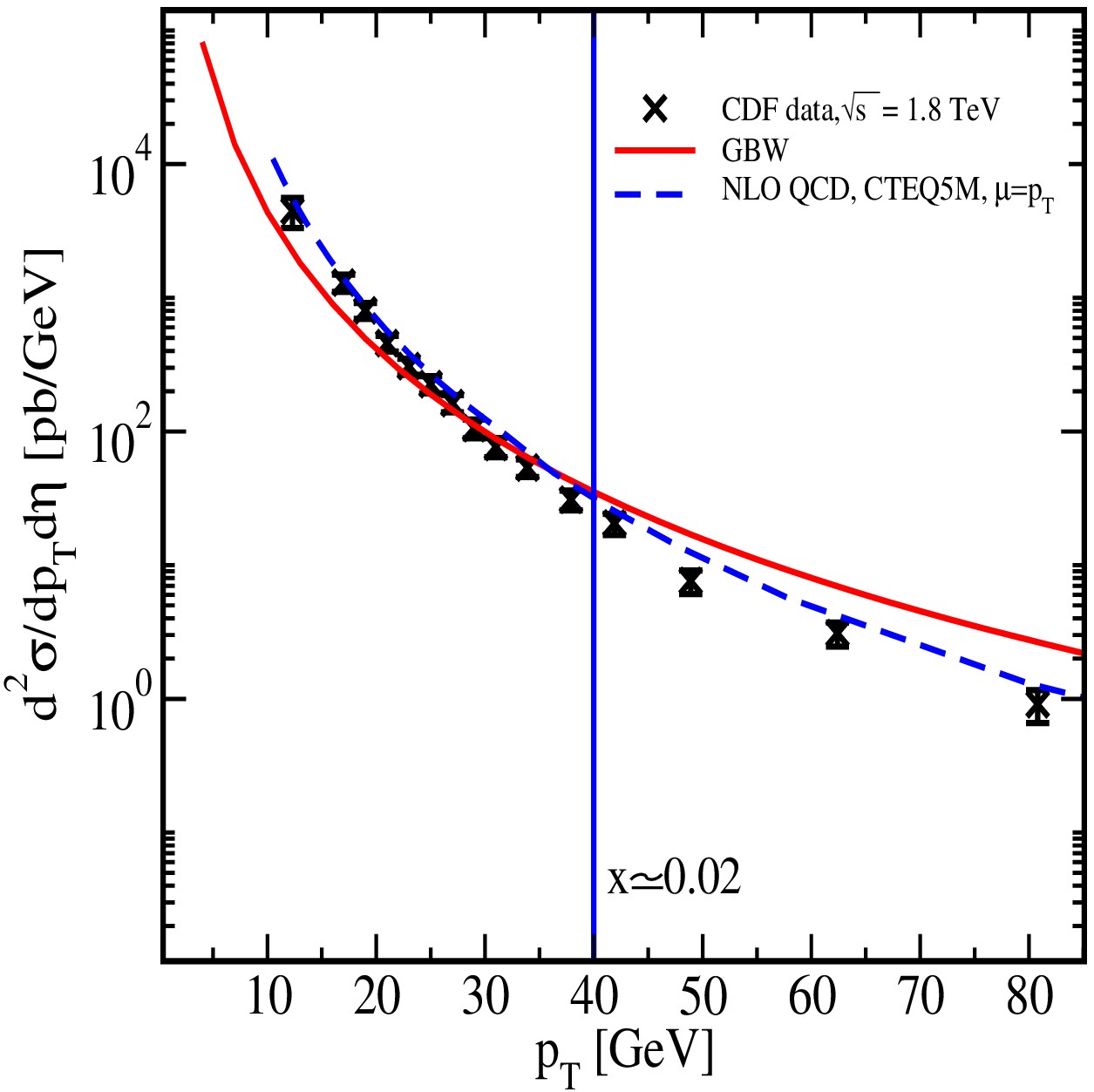}
            \caption{ Direct photon spectra obtained from the GBW dipole model at the RHIC and CDF energies for $pp$ collisions. We also show the NLO pQCD curve from the
authors of reference \cite{ppqcd} (given in table 3 of
Ref.~\cite{cdf1}) which used the CTEQ5M parton distribution functions with all scales set to $p_T$. Experimental data are from the PHENIX experiment \cite{rhicp} at $\eta=0$, and from the CDF experiment \cite{cdf1,cdf} at
$|\eta|<0.9$.  The error bars are the linear sum of the statistical
and systematic uncertainties.
\label{phon}}
\end{figure}
\begin{figure}[!t]
       \includegraphics[width=7 cm] {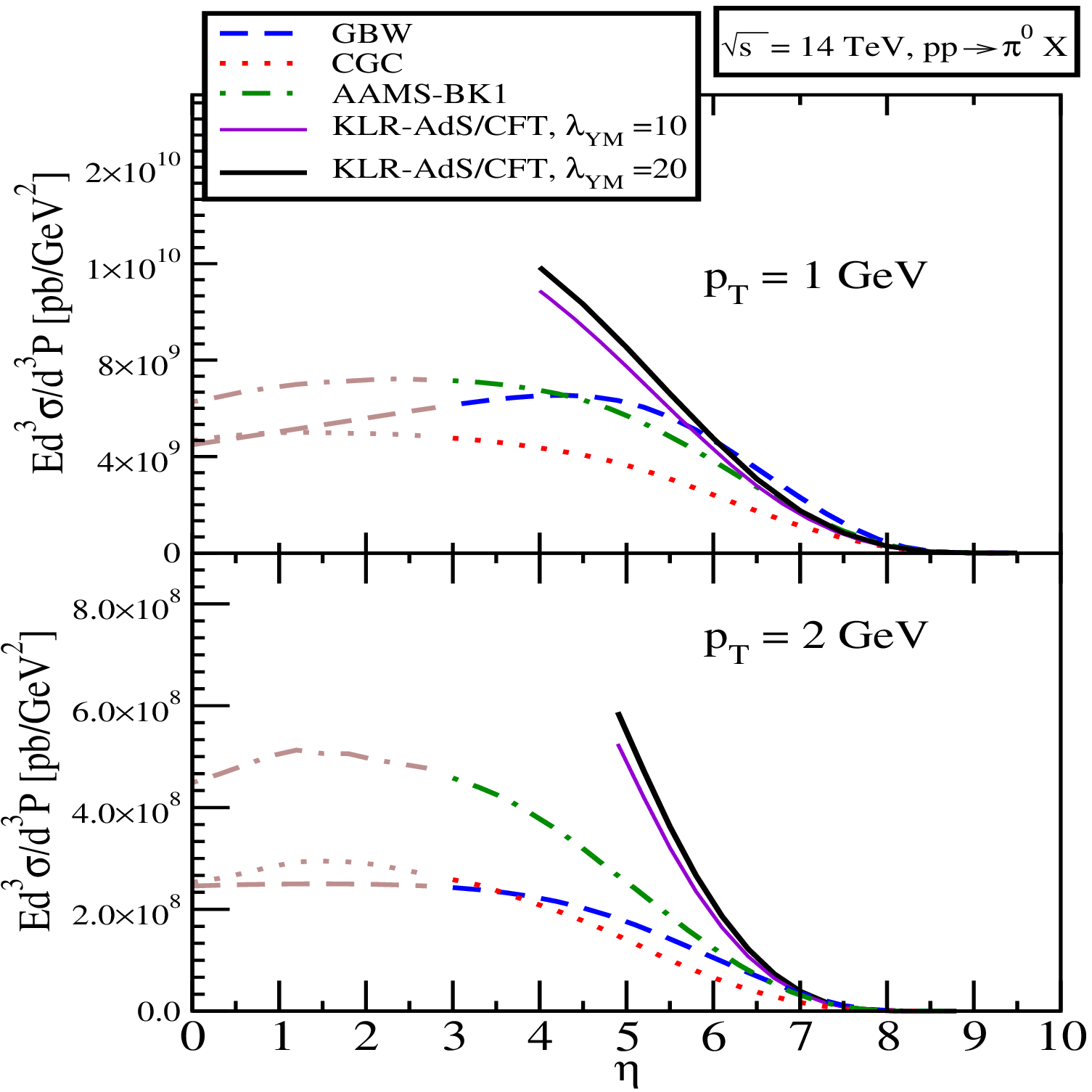}
       \includegraphics[width=7 cm] {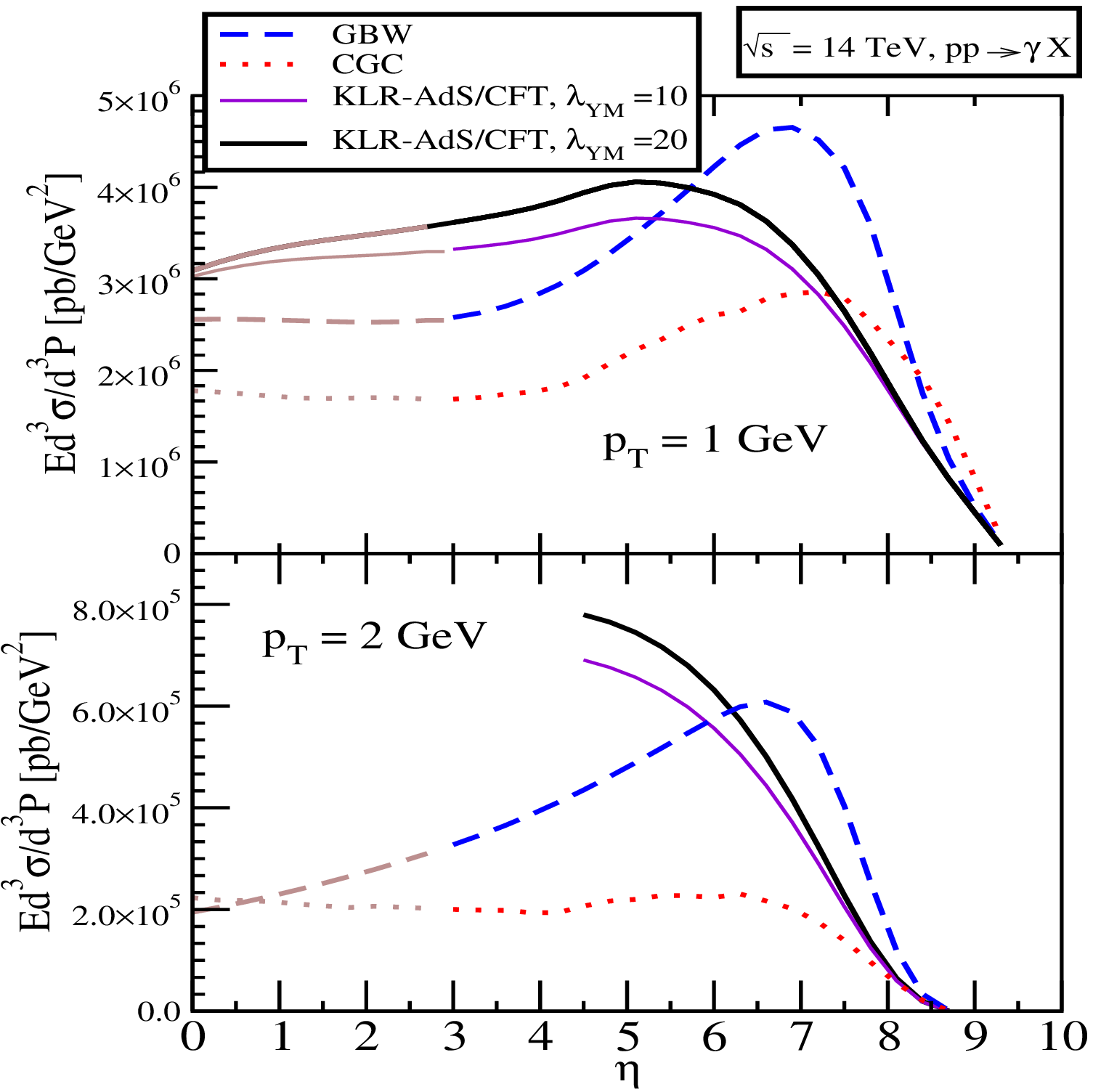}
        \caption{Invariant cross-section for pion (top) and direct photon (lower panel) production in $pp$ collisions at LHC as a
function of rapidity $\eta$ calculated with various color dipole models for various fixed
$p_T$. Our results for $pp$ collisions is less reliable at midrapidity (shown with brown color).  \label{lhc-e1}}
\end{figure}

In Fig.~\ref{phon}, we show direct photon
spectra obtained in our color-dipole approach Eq.~(\ref{con1}), at the RHIC \cite{rhicp}
($\sqrt{s}=200$ GeV) and CDF ($\sqrt{s}=1.8$ TeV) energy \cite{cdf,cdf1}.
Again, we should warn that our results at high $p_T$ for lower
energies like RHIC and CDF are less reliable since $x_2>0.01$ which is beyond the limit of applicability of the color-dipole
light-cone factorization scheme. Nevertheless, the agreement of our
results with available data for both hadron and photon production at
RHIC and CDF energies is rather satisfactory for $x\le 0.01$.  As a
comparison, in Fig.~\ref{phon}, we also show the NLO pQCD curve for
CDF energy \cite{ppqcd}. The predictions for direct photon spectra at LHC
energies in $pp$ collisions within various color-dipole models can be found in Ref.~\cite{amir1}.

In Fig.~\ref{lhc-e1}, the differential cross-section of pion $\pi^0$
(top panel) and direct photon $\gamma$ (lower panel) production at LHC are plotted versus rapidity at
fixed transverse momenta $p_T =1$ and $2$ GeV within various
color-dipole models. It is seen that the discrepancies among various
saturation color dipole model results can be about a factor of $2-3$ at
moderate rapidities.  At the kinematic limit, i.e. at very forward
rapidities and higher $p_T$ where the differential cross-section
approaches zero, kinematic constraints limit the parton phase space
and saturation effects become less important. This is seen in
Fig.~\ref{lhc-e1} where as we approach very forward rapidities at the kinematic limit, the
discrepancies among various saturation models shrink, and the invariant
cross-section identically approaches zero.  Notice that for hadron
production in the master Eq.~(\ref{pp1}), the light-cone momentum fraction $x\equiv\frac{x_2}{z}$
(where $0<z<1$ is the fragmentation fraction) enters the gluon
radiation cross-section and therefore the color dipole
cross-section, while in the case of direct photon production
Eq.~(\ref{con1}), we have $x\equiv x_2$. Therefore, the applicability of the KLR-AdS/CFT model
which is valid for $x< 6\times 10^{-5}$ (and $p_T^2<2.5~\text{GeV}^2$), can be extended for direct
photon production to lower rapidities compared to the case of
hadrons. It is seen from Fig.~\ref{lhc-e1} that for both hadron and
photon production, away from the kinematic limit, at not very large $\eta$ and
$p_T$, a color-dipole model with larger saturation scale leads to a stronger 
peak at forward rapidity (having in mind that the saturation scale is a dynamical function of $x$, see Fig.~\ref{f:sat}).

\begin{figure}[!t]
       \includegraphics[width=7 cm] {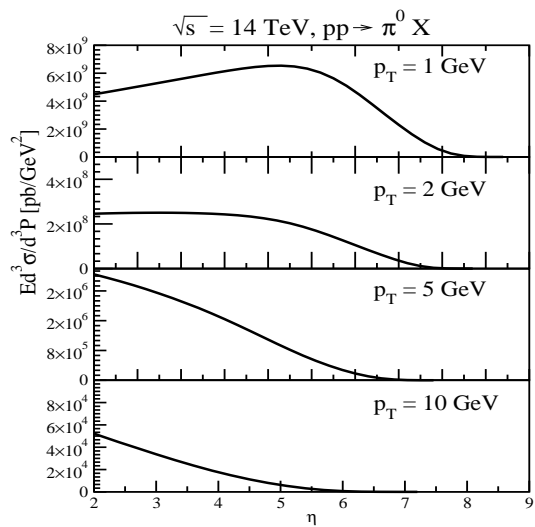}
            \caption{ Invariant cross-section for pion production in $pp$ collisions at LHC as a
function of rapidity $\eta$ calculated with the GBW color dipole model for various fixed
$p_T$. \label{lhc-e2}}
\end{figure}
\begin{figure}[!t]
       \includegraphics[width=7 cm] {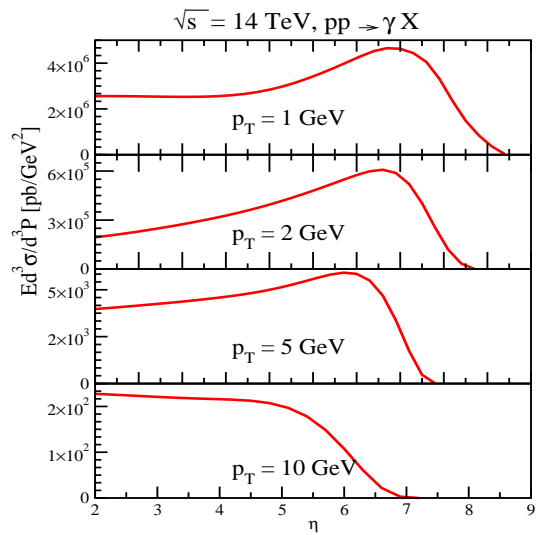}
            \caption{ Invariant cross-section for direct photon production in $pp$ collisions at LHC as a
function of rapidity $\eta$ calculated with the GBW color dipole model for various fixed
$p_T$. \label{lhc-e3}}
\end{figure}

In Fig.~\ref{lhc-e1}, it is seen a peculiar enhancement of the photon
production rate at forward rapidities. This feature is more obvious in
Figs.~\ref{lhc-e2} and \ref{lhc-e3} where we plot the differential
cross-section of pion and direct photon production at LHC as a function of rapidity at fixed transverse momenta $p_T
=1,2,5$ and $10$ GeV within the GBW model. It is obvious that the
invariant cross-sections have a peak at forward rapidity. However,
compared to pions, the peak of the differential cross-section for direct
photon production persists at larger $p_T$. It seems that several
mechanisms are at work here in different kinematic regions.  Looking
again at Fig.~\ref{lhc-e1} it is obvious that in the case of direct
photons when the saturation scale is smaller (the CGC model) at higher
transverse momentum $p_T=2$ GeV, the peak disappears and will be
replaced by a plateau. However, in the case of pion production, the
peak is less pronounced even in the presence of a large saturation scale, see
Figs.~\ref{lhc-e1} and \ref{lhc-e2}.  Moreover, photons are
radiated by the electric current of the projectile quarks, which
mostly stay in the fragmentation region of the beam, and tend to form
a peak at forward rapidities. However, at very large $p_T$ and $\eta$, the kinematic limit 
pushes photon radiation to more central rapidities and the peak at
forward rapidities will be replaced by a kind of plateau at central
rapidities.  At the same time, gluons are radiated via nonabelian
mechanisms by the color current across the whole rapidity interval and
tend to form a plateau at midrapidity.

\begin{figure}[!t]
       \includegraphics[width=7 cm] {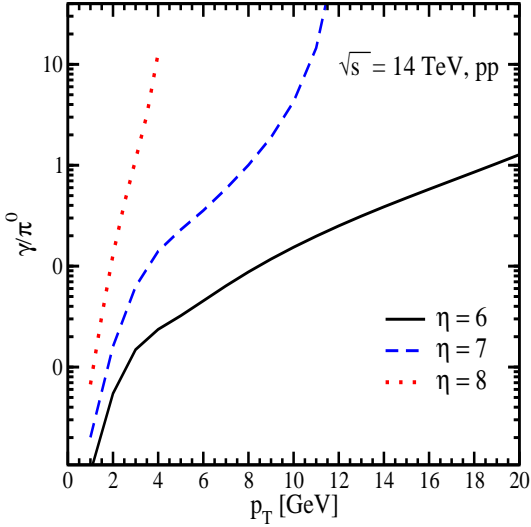}
            \caption{ The ratio of photon/pion production in $pp$ collisions at LHC as a
function of $p_T$ calculated with the GBW color dipole model.\label{lhc-r1}}
\end{figure}

Another interesting difference between direct photon and hadron
production is that direct photon production extends to higher
rapidities for a fixed $p_T$, see Figs.~\ref{lhc-e2}
and \ref{lhc-e3}. This is more obvious in Fig.~\ref{lhc-r1} where we
show the photon/pion ratio $\gamma/\pi^0$ as a function of $p_T$ at
various rapidities within the GBW model and $pp$ collisions. The ratio
$\gamma/\pi^0$ can be as big as $10-20$ at very forward
rapidities $\eta=8-7$ at LHC energy. Note that
suppression of hadrons at very forward rapidity also ensures significant
suppression of radiative decays of those hadrons. Therefore, direct
photon production at forward rapidities should be a rather clean signal.

\begin{figure}[!t]
       \includegraphics[width=7 cm] {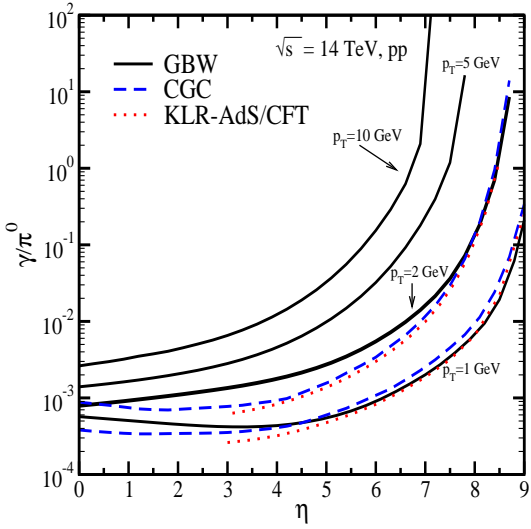}
            \caption{ The ratio of photon/pion production in $pp$ collisions at LHC as a
function of rapidity $\eta$ calculated with various color dipole models.\label{lhc-r2} }
\end{figure}

In Fig.~\ref{lhc-r2}, we show the ratio of photon/pion production as a function
of rapidity in $pp$ collision at LHC for various fixed $p_T$
within different saturation models.  Direct photons can only be 
radiated from quarks, while hadrons can be produced by both gluons and
quarks. At the LHC energy at midrapidity gluons dominate. Therefore
the photon/pion ratio is significantly reduced toward midrapidity. However, at
very forward rapidity, valence quarks become important and the
photon/pion ratio rises. Moreover, at high $p_T$ again valence quarks
becomes important and we have a sharp rise of the photon/pion ratio, see
Fig.~\ref{lhc-r2}. A similar behavior has also been reported in a different approach \cite{jam}.


\begin{figure}[!t]
       \includegraphics[width=8 cm] {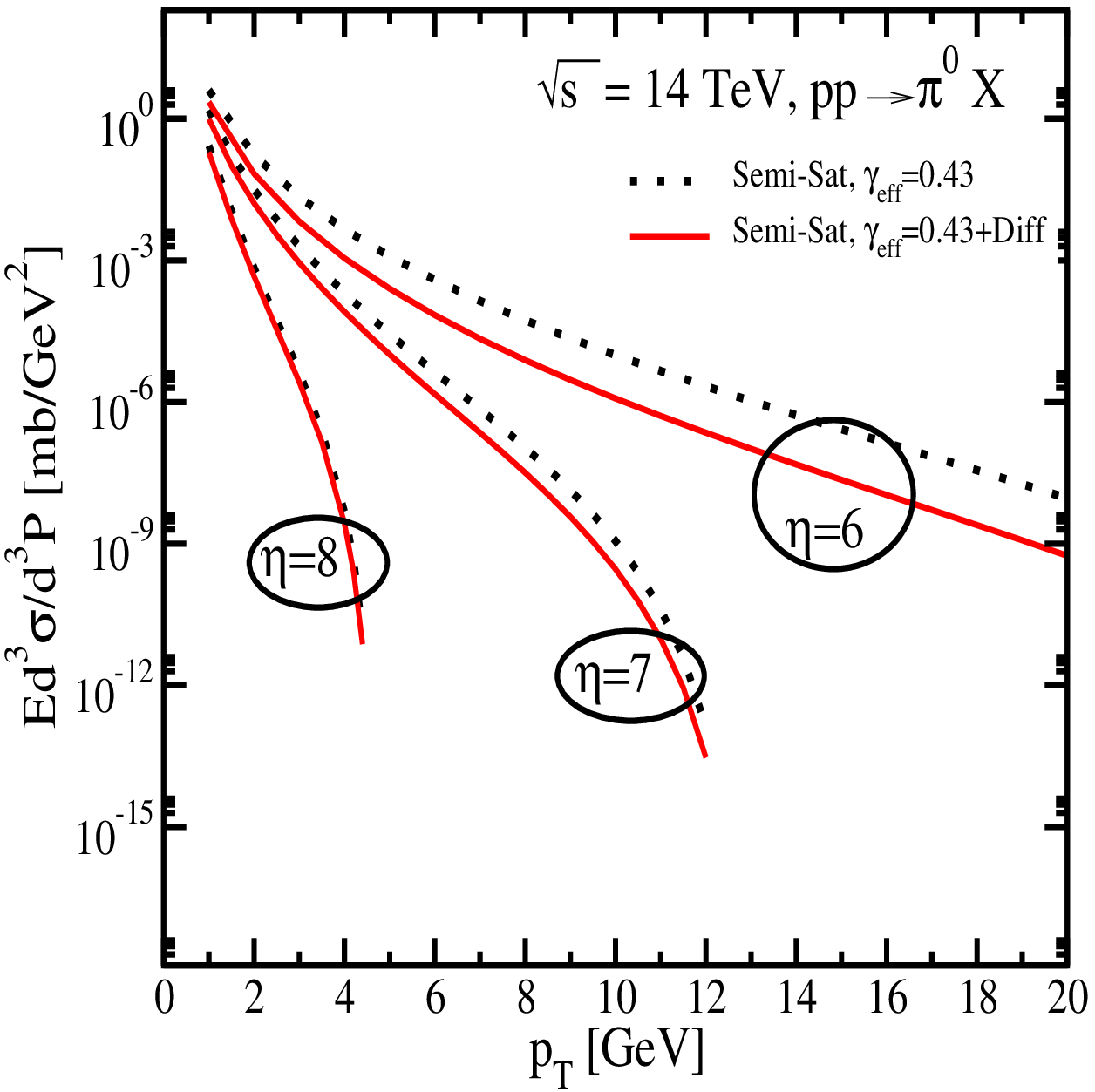}
       \includegraphics[width=8 cm] {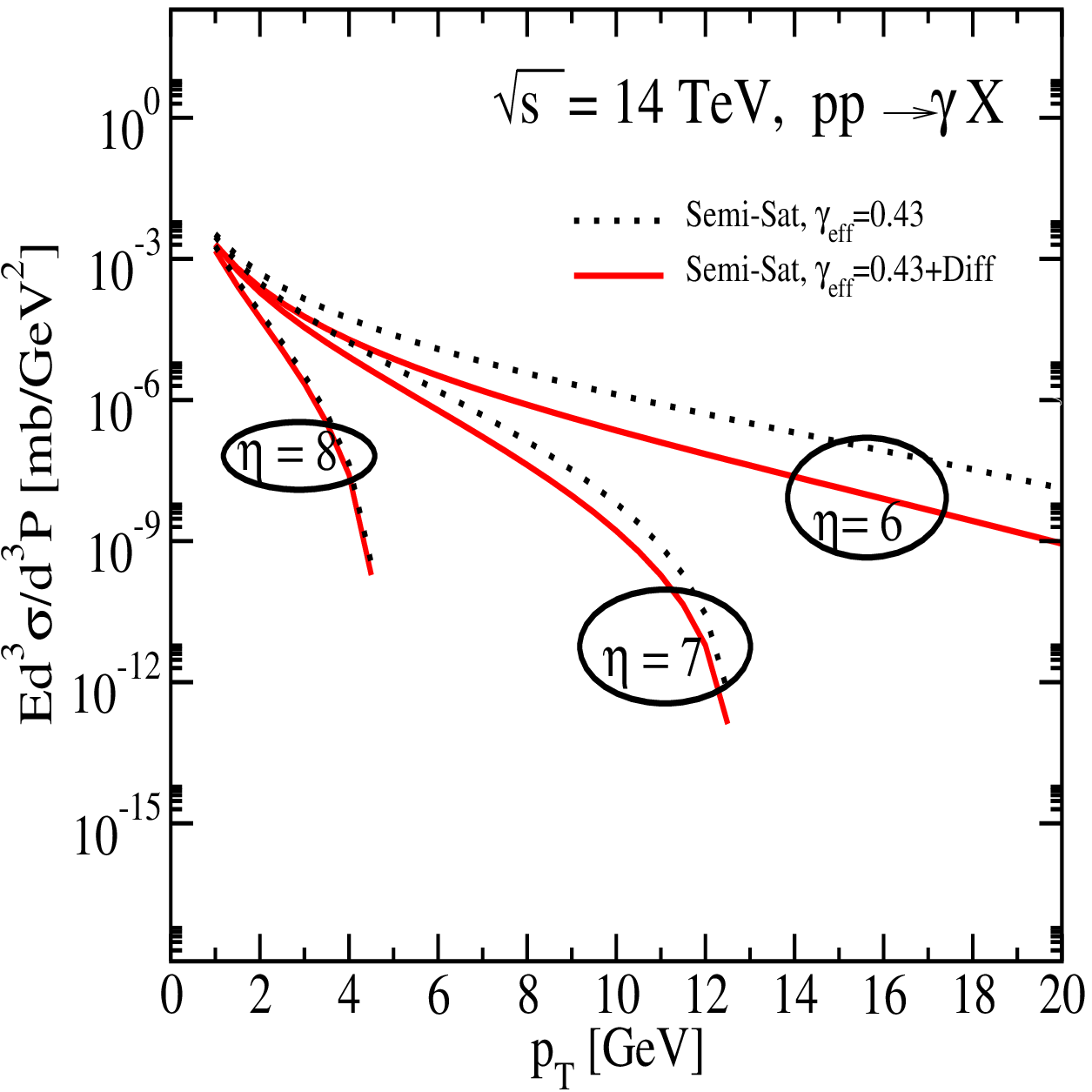} \caption{Pion (top)
       and direct photon (lower) spectra obtained from the Semi-Sat
       dipole model with two different effective anomalous dimension $\gamma_{eff}$  at LHC and forward rapidities in $pp$
       collisions. \label{lhc-s1} }
\end{figure}
\begin{figure}[!t]
       \includegraphics[width=8 cm] {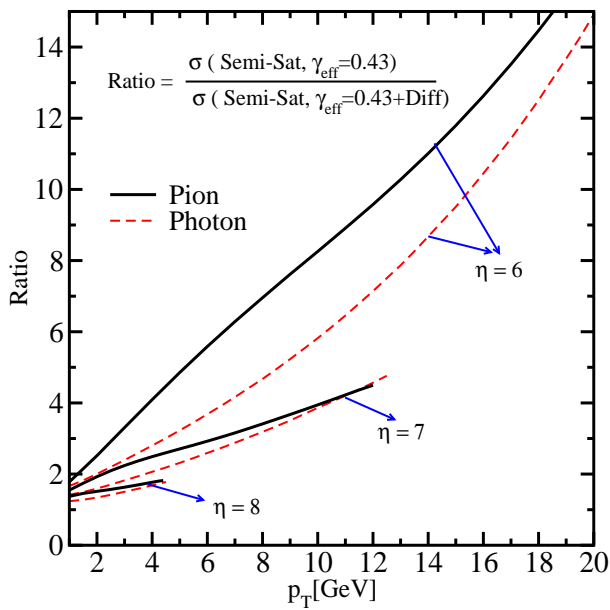}
            \caption{ The ratio of two invariant cross-sections obtained by taking two different effective anomalous dimension $\gamma_{eff}$ in the Semi-Sat
       dipole model for pion and direct photon production at LHC energy ($\sqrt{s}=14$ TeV) in $pp$ collisions at various rapidities.   \label{lhc-s2}}
\end{figure}

In order to understand the relative importance of saturation effects
at various rapidities, we employ the Semi-Sat model.  In Fig.~\ref{lhc-s1} we show, the differential
cross-section of pion and photon production at LHC, calculated once
with diffusion term and once without, i.e. $\gamma_{eff}=0.43$. We
recall that the Semi-Sat model in the presence of the diffusion term
describes $F_2$-data at HERA, see Sec.VI-C.  
In Fig.~\ref{lhc-s1} we show that at
forward rapidities, the diffusion term in the anomalous dimension is
not important, since it gives similar results as with a fixed
$\gamma_{eff}=0.43$. The preferred value of anomalous dimension $1-\gamma_{eff}=0.57$ at very forward rapidities
is close to the one predicted from the BK equation \cite{bkk}.  This is more obvious in Fig.~\ref{lhc-s2} where we
show the ratio of the two cross-sections for both pions and direct
photons. It is well known that the saturation effects start being
essential when the anomalous dimension reaches the value $\gamma_{cr}
= 1-\gamma_{eff}=0.37$ which is the case for forward rapidities (see Refs.~\cite{Gribov:1984tu1,GAMMA,lbk}). This
indicates that direct photon and hadron production at different
rapidities at LHC are rather sensitive to saturation. 


\section{ Cronin effect and nuclear modification factor }

The nuclear modification (Cronin) factor $R_{pA}$ is defined as ratio of $pA$ to
$pp$ cross-sections normalized to the average number of binary nucleon
collisions,
\begin{equation}
R_{pA}=\frac{ \frac{d \sigma^{pA\to h+X}}{dy d^2 p_T}}{\langle N_{binary}\rangle\frac{d \sigma^{pp\to h+X}}{dy d^2 p_T}}.
\end{equation}
$\langle N_{binary}\rangle$, the average number of geometrical
binary collisions, is calculated according to the Glauber model \cite{gl} for different centralities. 
 
Two very different mechanisms have been proposed to explain Cronin
enhancement (or suppression) in $pA$ collisions: a) initial-state
effects \cite{cron,boris1,break,cro-cgc1,cro-cgc2} due to a broadening of the parton transverse
momentum in the initial-state. Here the fragmentation of hard partons
is assumed to occur outside the cold medium. b) final-state effects
\cite{fin1} due to the recombination of soft and shower partons in the
final-state.

In our approach, the Cronin effect originates from initial-state
broadening of the transverse momentum of a projectile parton
interacting coherently with a nuclear medium. The invariant
cross-section of hadron and direct photon production in $pA$
collisions can be obtained via the light-cone color-dipole
factorization scheme defined in Eqs.~(\ref{pp1},\ref{con1}).

\subsection{Gluon shadowing}

In the infinite momentum frame, the gluon clouds of nucleons which
have the same impact parameter overlap at small Bjorken-x in the
longitudinal direction. This allows gluons 
which originate from different
nucleons to fuse, corresponding to a nonlinear term in the evolution
equation which suppresses gluon production, and a precocious onset of the saturation effects for heavy
nuclei. This is called gluon shadowing. The same effect, looks different in the rest frame of the nucleus, the gluon shadowing correction can be 
calculated as Landau-Pomeranchuk effect, namely the
 suppression of bremsstrahlung by interference
of radiation from different scattering centers. This mechanism 
requires a sufficiently long coherence time for
radiation, a condition equivalent to requiring 
a small Bjorken-x in the parton model.

The question if gluon shadowing is an intrinsically leading
twist effect \cite{d1-n} or is suppressed by power of $Q^2$ and is 
due to higher twist/high parton density effects \cite{Gribov:1984tu1,Gribov:1984tu} is still
debatable. There has been several shadowing models which consider only
leading twist shadowing, e. g., including shadowing effects in the
non-perturbative initial conditions which are then evolved with 
leading twist DGLAP equations \cite{d2-n}. 
Modifications of this leading twist picture to include Mueller-Qiu type
non-linear contributions has been studied in Ref.~\cite{d3-n}.

In our approach, nuclear shadowing for gluons is calculated from
shadowing of the $|q \bar q g\rangle$ Fock component of a
longitudinally polarized photon. Unlike transverse photons, all $q
\bar q$ dipoles from longitudinal photons have size $1/Q^2$ and the
double-scattering term vanishes like $1/Q^4$. The leading-twist
contribution for the shadowing of the longitudinal photons arises,
therefore, 
from
the $|q \bar q g\rangle$ Fock component. While the
$q \bar q$ separation is of order $1/Q^2$, the gluon can propagate
relatively far from the $q\bar q$-pair. After gluon radiation $q\bar
q$ is in a color octet state, consequently the $q \bar q g$ system
appears as $gg$ dipole. The shadowing correction to the longitudinal
cross-section is then directly related to gluon shadowing. The gluon shadowing ratio is defined as the ratio of the gluon densities in 
a nucleus and a nucleon \cite{kst2,shad1}:

\begin{equation}
R_G(x,Q^2, b)= \frac{G_A(x,Q^2,b)}{A G_N(x,Q^2)}\approx 1-\frac{\Delta\sigma_{L}^{\gamma A}[q\bar q g](x,Q^2,b)}{A\sigma_{L}^{\gamma p}(x,Q^2)}, \label{rss}
\end{equation}
where $\Delta\sigma_{L}^{\gamma A}[q\bar q g]$ is the inelastic
correction to the longitudinal photoabsorption cross-section
$\sigma_{L}^{\gamma A}$ due to the creation of a $|q \bar q g\rangle$
Fock component.  The details for the calculation of the suppression factor
$R_G$ can be found in Refs.~\cite{kst2,shad1,shad2}. For a proton target, we have $R_G=1$ by
construction.

At high energy the $q\bar{q}$ dipole cross-section is also subject to
the multi-pomeron fusion effects in a nuclear medium. These effects
are missed in the eikonal formulas
Eqs.~(\ref{eik2},\ref{eik0},\ref{eik00},\ref{eik1}) where the
variation of the transverse size of the $q\bar{q}$ Fock component
while propagating and interacting with a medium was not taken into
account.  Consequently higher Fock components were summed up without
incorporating gluon shadowing. One should note that the multiple
parton interactions that lead to gluon shadowing are also the source
of gluon saturation. In order to avoid double counting, we calculate
the nuclear shadowing effect within the same color-dipole
formulation. The authors of Refs.~\cite{kst2,shad1} have performed
such a calculation by numerically solving the $q\bar{q}$ dipole
evolution equations in a medium by light-cone Green function
techniques and confronted DIS data for nuclei. Following
Refs.~\cite{boris1,break,kst2,shad1,shad2} one can effectively
incorporate gluon shadowing due to the nuclear medium by modifying the
cross-section of the $q\bar{q}$ dipole interacting with a nucleus
target at impact parameter $b$ by the following replacement
\begin{equation}
\sigma_{q\bar{q}}(r,x)\to R_G(x, Q^2, b)\times \sigma_{q\bar{q}}(r,x),
\end{equation}
in the exponent of
Eqs.~(\ref{eik2},\ref{eik0},\ref{eik00},\ref{eik1}). Therefore, by
means of $R_G$ and the dipole cross-section on a {\em nucleon} target,
one can effectively define the $q\bar q$ dipole cross-section for a
{\em nucleus} target by using Glauber theory, i.e. via simple
eikonalization of the $q \bar q$-nucleon cross-section modified by the
suppression factor $R_G$. In this way, we relate the nuclear gluon
shadowing to the gluon saturation which can be then read off from the
constructed dipole-nucleus forward amplitude. However, the question if
the parton saturation provides a precise microscopic understanding of
shadowing is an open question and out of scope of this
paper. One should also note that although the shadowing factor $R_G$ improves the eikonal
approximation, it is not apparently a solution of the non-linear BK evolution equation.

In the CGC picture, the dipole-nucleus amplitude has
the same functional form as the dipole-nucleon amplitude. The only
difference is the saturation scale. The $A$-dependence of
dipole-nucleus amplitude enters through the saturation scale
$Q_{sA}^2\approx Q^2_{s} A^{1/3}$ where $A$ is the effective mass
number of the nucleus in a given centrality and depends on the impact
parameter. Our approach is different but is not in contradiction with the CGC
picture at the saturation boundary. In order to see this, let us
assume that in spirit of the CGC picture one can write the forward
dipole-nucleus amplitude in the following form (we use the GBW form
for simplicity),
\begin{equation}
\mathcal{N}_{q\bar{q}}^{A}(r,x)=1-e^{-(rQ_{sA}(x))^{2}/4}. \label{gbw1-a}
\end{equation}
By comparing the above equation with Eq.~(\ref{eik0}) and assuming that there is no correlation between dipole amplitude and the nuclear
thickness, one can immediately read off the effective saturation scale in  Eq.~(\ref{eik0}) close to the saturation boundary,  
\begin{eqnarray}
Q_{sA}^{2}(x,b)&=& 2\sigma_0 R_G(x,Q^2)\frac{\sigma_{q\bar{q}}^{N}(r,x)}{r^2} T_A(b), \label{qsa}\\
&\approx& 2\sigma_0 R_G(x, Q^2) Q_{sN}^2(x)T_A(b),  \
\end{eqnarray}
where in the second line we rely on the small-$r$ approximation of the
dipole cross-section (valid for a large $T_A(b)$) and use the fact that
$R_G\to 1$ at $r\to 0$ since $Q^2\sim 1/r^2\to \infty$. Therefore, the
square of saturation scale $Q_{sA}^2$ in our approach is approximately
proportional to $A^{1/3}$ since $T_A(b)\sim A^{1/3}$, in agreement with
the basic idea of saturation and the CGC picture \cite{Gribov:1984tu1,Gribov:1984tu,Iancu:2003xm,mul}. 
Let us repeat the above steps in a slightly different way. 
The dipole-nucleon cross-section at small dipole
size $r$ can be related to the gluon distribution $xG(x, Q^2)$ in the nucleon \cite{small-r},
\begin{equation}
\sigma_{q\bar{q}}^{N}(r,x)=\frac{\pi^2}{3}\alpha_s(1/r^2)xG(x,1/r^2)r^2.   
\end{equation}
By plugging the above expression into Eq.~(\ref{qsa}) we obtain,
\begin{equation}
Q_{sA}^{2}(x,b)=\frac{2\pi^2}{3}R_G(x, Q^2)\alpha_s(1/r^2)xG(x,1/r^2)T_A(b),\label{qsa-2}
\end{equation}
where the typical value of dipole size can be related to the saturation scale $Q^2_{sA}\sim 1/r^2$. This is remarkably similar to the  
saturation scale proposed by  by Kharzeev, Levin and Nardi (KLN model) \cite{KLN},
\begin{equation}
Q_{gA}^{2}(x,b)=\frac{3\pi^2}{2}\alpha_s(Q^2_{gA})xG(x,Q^2_{gA})\rho_{part}^{A}(b), \label{qsa-3}
\end{equation} 
where for $pA$ collisions the density of participants is
$\rho_{part}^{A}(b)=T_A(b)$. Note that Eq.~(\ref{qsa-3}) gives the
saturation scale for gluons and it is different from the saturation
scale for quarks Eq.~(\ref{qsa-2}) by a Casimir factor $9/4$. The KLN
model Eq.~(\ref{qsa-3}) gives a good description of hadron
multiplicities in heavy ion collisions at RHIC \cite{KLN}.  The main
difference between our model Eq.~(\ref{qsa-2}) and the KLN model
Eq.~(\ref{qsa-3}) is the shadowing factor $R_G(x,Q^2)$ which takes
into account approximately multi-pomeron fusion effects in a nuclear
medium beyond the eikonal approximation, see also Ref.~\cite{gng}.

In the limit of strong shadowing at very small $Q^2$, the gluon ratio
Eq.~(\ref{rss}) has a simple form $R_G\approx \pi
R^2_A/(A\sigma_{eff})$ where $\sigma_{eff}$ is the effective
cross-section responsible for shadowing and $R_A$ is the nuclear
radius. Therefore, in our approach, deep inside saturation region, we have $R_G\to
1/T_A(b)$ and consequently the saturation scale $Q_{sA}$ becomes
independent of $A$.  This behavior has been also
predicted based on more sophisticated models indicating that the parton wave
functions of different nuclei become universal at high energies limit
\cite{mul2,KLN2}. Nevertheless, we expect that our approach based on an improved
eikonal approximation will not be reliable at such an extreme limit and we use
our formulation only at midrapidity for $pA$ collisions at RHIC and
LHC energies.

As we argued above, in
principle one may construct the dipole-nucleus amplitude via the
dipole-nucleon amplitude supplemented with the $A$-dependent saturation
scale. However, it is not a priori obvious whether such a model with parameters fitted to the available DIS data on proton target is also 
able to describe the DIS data on nucleus target at small-$x$ without having to change the parameters of the model (the issue of sensitivity of the model parameters obtained from a fit to data in $\chi^2$ analysis),  see Ref.~\cite{her}. 
In our approach we use the same dipole-nucleus
cross-section which gives a good description of HERA data to calculate
the cross-section in $pA$ reactions. We stress again that the shadowing factor $R_G(x,Q^2)$ is not a free
parameter in our formalism but it is calculated via
Eq.~(\ref{rss}). Such a shadowing factor is needed in order to
describe the DIS data off nuclei
\cite{boris1,break,kst2,shad1,shad2}. Therefore the suppression obtained
as a result of the inclusion of the shadowing factor $R_G$ (which
depends on kinematics) is not arbitrary.

\begin{figure}[!t]
       \includegraphics[width=7 cm] {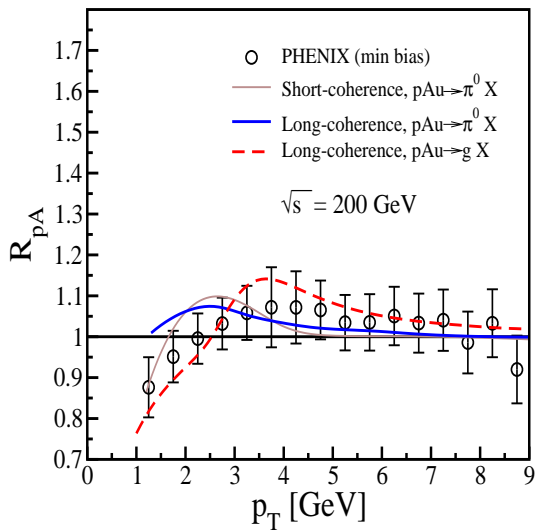} \caption{Nuclear modification factor $R_{pA}$
       for pion production at RHIC energy in minimum bias proton-gold collisions at
       midrapidity within the long-coherence length scheme presented in this
       paper. Note that at RHIC energy we are in the transition region
       between the short and long-coherence length limit. For comparison, we
       also show the Cronin curve obtained in the short-coherence length scheme \cite{me-croin}.   The Cronin ratio for
       gluon production in the long-coherence length scheme is also shown. The
       GBW saturation model is used for all curves. The experimental data are from \cite{rhic2006}.  \label{cr1}}
\end{figure}

\subsection{Numerical results for $pA$ collisions}

For the calculation of cross-sections for $pA$ collisions, we use the
same PDFs and FFs as for $pp$ collisions. Furthermore, we use 
a Woods-Saxon nuclear profile for $T_A(b)$.  We again stress that
similar to the calculation for $pp$ collisions, here again we have no
free parameters to adjust. In Fig.~\ref{cr1}, we show 
$R_{dAu}$ for $\pi^0$ production at RHIC in minimum bias proton-gold collisions. The experimental data in Fig.~\ref{cr1} are
from PHENIX \cite{rhic2006}. As we already mentioned, for RHIC energy at
midrapidity and moderate $p_T$, the coherence length defined via
Eq.~(\ref{g-2g}) is about $l_c\sim 5-6$ fm which is comparable to
the nuclear radius. Therefore, we are in the transition region
between the regimes of long and short coherence length. Calculations in
such a region are most complicated. In Fig.~\ref{cr1}, we show the
theoretical curves calculated in the two extreme cases of short- and
long-coherence length. The curve for the short-coherence length in Fig.~\ref{cr1} is based on an
improved pQCD calculation taken from Ref.~\cite{me-croin}. We used the AKK08 for FFs, MSTW2008
for PDFs and the GBW model for the color dipole cross-section. One should also note
that the color dipole cross-section is fitted to the DIS data for $x_2
\le 0.01$. Therefore, our results at high $p_T$ for RHIC are
less reliable. At RHIC and LHC energies at
midrapidity, gluons are mostly responsible for pion production. We
also show in Fig.~\ref{cr1} the Cronin ratio for gluon production in
proton-gold collisions. It is seen that fragmentation processes 
distort the gluonic Cronin enhancement and shift the Cronin peak to a lower
$p_T$. 

\begin{figure}[!t]
              \includegraphics[width=7 cm] {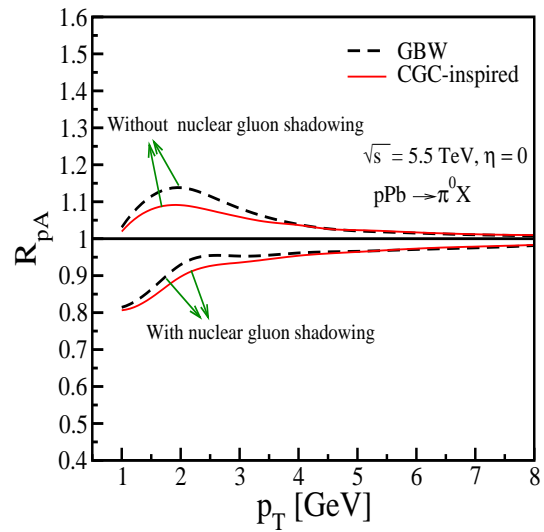}
              \caption{Nuclear modification factor
              $R_{pA}$ for pion production at the LHC energy
              $\sqrt{s}=5.5$ TeV at midrapidity in minimum bias proton-lead
              collisions within the CGC and GBW color dipole
              models. We show results with and without inclusion of the nuclear gluon shadowing factor $R_G(x, Q^2, b)$.
              \label{clhc1}}
\end{figure}

\begin{figure}[!t]
              \includegraphics[width=7 cm] {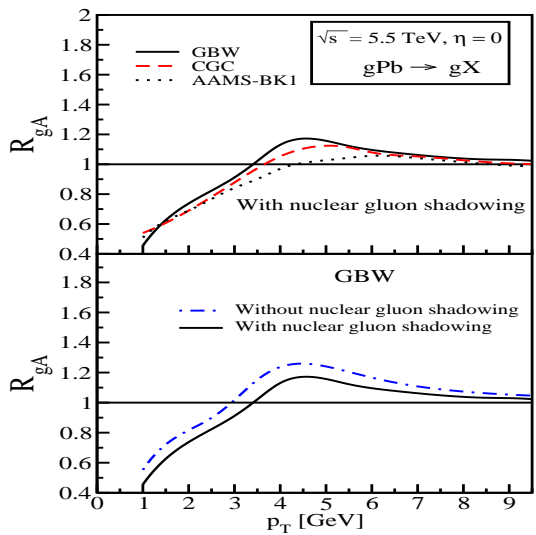}
              \caption{Same as Fig.~\ref{clhc1} for gluon
              production. In all curves in the upper panel the nuclear
              gluon shadowing factor $R_G$ (defined via Eq.~(\ref{rss}))
              is incorporated. Lower panel: gluon shadowing effects at
              LHC for the GBW model. \label{clhc2}}
\end{figure}

In Fig.~\ref{clhc1}, we show our prediction for the 
nuclear modification factor $R_{pA}$ for $\pi^0$ production at LHC at
midrapidity in minimum bias $pA$ collisions within two very different saturation
models, namely GBW and CGC. We also show the effect of nuclear gluon
shadowing.  It is seen that the Cronin enhancement will be replaced
with moderate suppression in all saturation models considered in this
paper due to nuclear gluon shadowing.  It is obvious that a $q\bar q$-proton dipole model with a bigger
saturation scale leads to a larger Cronin enhancement and works
against the nuclear shadowing suppression. This effect has also been shown in
Ref.~\cite{sa-sa}. Note that the source of both saturation and
shadowing is parton multiple interaction. However, a larger
saturation scale leads to a stronger broadening of transverse momentum
of the projectile partons and consequently it works against
shadowing. This is more obvious in Fig.~(\ref{clhc2}) (uppor panel)
where we plotted the Cronin ratio for gluons production at the LHC
energy within various saturation color dipole models.  In
Fig.~(\ref{clhc2}) (lower panel) we show effect of nuclear gluon shadowing
within the GBW color dipole model. It is seen that both
shadowing and saturation effects are important at LHC 
in $pA$ collisions and give rise to a rather sizable effect in the
nuclear modification factor $R_{pA}$.
 
\begin{figure}[!t]
              \includegraphics[width=7 cm] {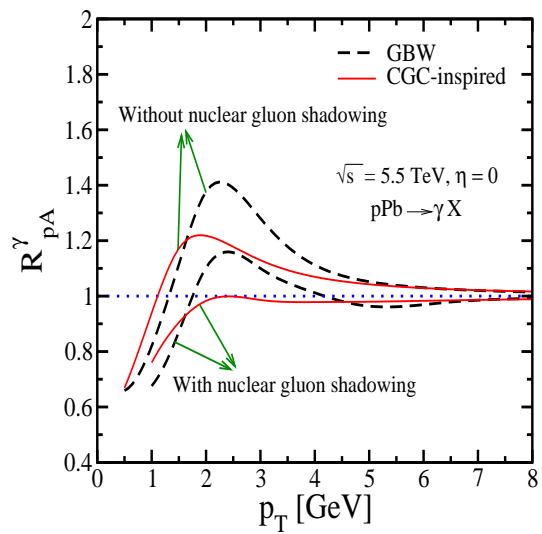}
              \caption{Same as Fig.~\ref{clhc1} for direct photon
              production. For comparison, we also show the results
              with and without inclusion of nuclear gluon shadowing
              effects introduced via Eq.~(\ref{rss}).\label{clhc3} }
\end{figure}

In Fig.~\ref{clhc3}, We show our prediction for the nuclear modification factor $R_{pA}^{\gamma}$ for
direct photon production at LHC at midrapidity in minimum bias $pA$ collisions for two models with different
saturation scale. In order to demonstrate the importance of nuclear gluon shadowing
effects, we have also plotted the curves without nuclear gluon shadowing. In comparison to pion production, the Cronin
enhancement for direct photon production seems stronger and survives within the
GBW color-dipole model which has a bigger saturation scale, even after
the inclusion of nuclear gluon shadowing suppression effects. Similar to pion
production, the Cronin enhancement for direct photon production is
bigger in a model with a larger saturation scale.  
Within the CGC model both pion and direct
photon enhancement at RHIC will be replaced by suppression at LHC.


In a similar approach, Kopeliovich {\em et al.} \cite{boris1} have shown that the Cronin enhancement will survive
at LHC at midrapidity though reduced compared to RHIC. Here, our finding is
different. This is due to the fact that we improved the earlier calculation
in several ways including: using updated PDFs, FFs and color-dipole
cross-sections, incorporating gluon radiation from
the quark projectile (i.e. $qN\to qgX$, see Eq.~(\ref{pp1})), using
recently upgraded shadowing suppression factor $R_g$ \cite{shad2} and using a running strong coupling. Nevertheless, in both
approaches the Cronin ratio $R_{pA}$ is still very small, less than
$20\%$ at midrapidity at LHC.  Kharzeev {\em et al.} \cite{cro-cgc1} have shown a
marked suppression for pions at midrapidity at LHC in $pA$ collisions based on the CGC scenario. This suppression is stronger than our
prediction. Certainly, LHC data should be able to decide between the different approaches and scenarios.

Notice that our prescription for both hadron and photon
production in $pA$ collisions is less reliable at very large $p_T$ and also forward
rapidities. This is due to the fact that at large $x_F$ (i.e.
$x_1\to 1$) one should properly incorporate energy conservation
since it puts an important constraint on particle
production \cite{break} . Nevertheless, we expect this effect to be negligible in our
kinematical region of interest. Note
also that the energy loss effects are subject to $x_1$-scaling and
less important for high-energy $pA$ collisions at moderate $p_T$, although
it might be important at lower energies \cite{break,boris1}.  A
more detailed study of the Cronin effect for direct-photon production at 
RHIC and LHC at forward rapidities and high $p_T$ will be presented elsewhere \cite{amir-co}.

\section{Conclusions and outlook}
In this paper, we investigated pion and direct photon production
within a unified color-dipole approach at high-energy $pp$ and $pA$
collisions and provided various predictions for the upcoming LHC experiments. The results of this paper can be summarized as follows:

\begin{itemize}

\item Both hadron and direct photon production strongly depend on the value of the anomalous dimension $\gamma_{eff}$ and are sensitive to gluon saturation effects 
at forward rapidities at LHC ($\sqrt{s}=14$ TeV). The difference between various saturation model predictions can be about a factor $2\div 3$. Note that all saturation models employed here are fitted to HERA data.

\item We showed that the ratio of photon/pion production at LHC ($\sqrt{s}=14$ TeV) at very forward rapidities in $pp$ collisions can be as big as $10-20$.  
Therefore, direct photons at very forward rapidities should be a rather clean observable and provide a
sensitive probe for saturation effects and small-x physics in general.

\item We showed that the rapidity distribution of pions and direct photons exhibit 
 some peculiar enhancement at forward rapidities which is more
 pronounced in the case of photon production. This peak is enhanced in models with a larger
 saturation scale at lower $p_T$. 

\item We investigated the relationship between saturation and shadowing effects in $pA$ collisions at LHC for both direct photon and
hadron production. We studied the role of initial-state broadening
of the transverse momentum distribution of a projectile parton propagating and
interacting coherently with a nuclear medium. We showed that a larger
saturation scale leads to a stronger transverse momentum broadening of
the projectile partons and consequently works against the
nuclear gluon shadowing suppression effects.  Our results show that the nuclear
modification factor $R_{pA}$ at LHC is sensitive to
both saturation and nuclear shadowing effects and it seems that a subtle
cancellation between these two effects leads to a rather small Cronin
ratio $R_{pA}$. We showed that the 
$\pi^0$ and direct photon $\gamma$ Cronin ratio $R_{pA}$ at the LHC is less than
$1$ within the CGC color dipole model.
However, in the case of direct photon production in $pA$ collisions, the Cronin enhancement can survive at the LHC energy within 
the GBW color-dipole model which has a larger
saturation scale. 

\end{itemize}


\begin{acknowledgments}
  
   We would like to thank Javier Albacete, Boris Kopeliovich, Genya Levin, Marco Stratmann and Heribert Weigert for useful discussions.

   We are thankful to Yuri Kovchegov for useful communication in connection with Ref.~\cite{c3-n}.

  A.R. acknowledges the financial support from the Alexander von Humboldt foundation, BMBF (Germany), Conicyt Programa
  Bicentenario PSD-91-2006, Fondecyt grants 1090312 (Chile).

\end{acknowledgments}  


\end{document}